\documentclass[aps,pra,preprint,groupedaddress]{revtex4-1}
\usepackage{amsbsy}
\usepackage{amssymb}
\usepackage{amsfonts}

\usepackage{fouridx}
\usepackage{graphicx}
\usepackage{hyperref}   
\usepackage[all]{hypcap} 

\usepackage{verbatim} 
\usepackage{amsmath}
\newcommand{\bs}{\boldsymbol}
\newcommand{\mz}{\mathbb{Z}}
\newcommand{\mr}{\mathbb{R}}
\newcommand{\ii}{{\rm i}}
\newcommand{\di}{\,{\rm d}}
\newcommand{\lims}{\stackrel{s \to 0}{\longrightarrow}}
\newcommand{\erfc}{{\rm erfc}}
\newcommand{\tabnote}{\caption}

\begin{document}
\title{Ewald method for polytropic potentials in arbitrary dimensionality}
\author{O. N. Osychenko}
\author{G. E. Astrakharchik}
\author{J. Boronat}
\affiliation{Departament de F\'{\i}sica i Enginyeria Nuclear, 
Universitat Polit\`ecnica de Catalunya, 
Campus Nord B4-B5, 08034 Barcelona, Spain}
\date{\today}

\begin{abstract}
The Ewald summation technique is generalised to power-law $1/|\bs{r}|^k$ potentials in three-, two- and one-dimensional geometries with explicit formulae for all the components of the sums. The cases of short-range, long-range and ``marginal'' interactions are treated separately. The jellium model, as a particular case of a charge-neutral system, is discussed and the explicit forms of the Ewald sums for such system are presented. A generalised form of the Ewald sums for a noncubic (nonsquare) simulation cell for three- (two-) dimensional geometry is obtained and its possible field of application is discussed. A procedure for the optimisation of the involved parameters in actual simulations is developed and an example of its application is presented.
\end{abstract}
\maketitle

\section{Introduction}
\label{sec:introduction}

The behaviour of many-body systems is often governed by the long-range
Coulomb potential between charged particles.  Numerical simulations of such
systems are usually performed by considering a finite number of particles
in a cell with periodic boundary conditions.  The correct estimation of the
potential energy in such systems requires of a summation over all images
created by the periodic boundary conditions. For long-range interaction
potentials such direct summation either converges slowly or it is
conditionally convergent, making its evaluation computationally cumbersome.
Instead, the performance of the calculation can be greatly improved by using
Ewald summation methods~\cite{ewald}. In these methods, the slowly
convergent tail of the sum in the potential energy is represented by a
rapidly convergent sum in momentum space. The method is named after Paul
Peter Ewald who in his pioneering work dated almost a century ago
calculated the electrostatic energy in ionic crystals (a detailed
derivation for the Ewald sums for the Coulomb potential can be found in the
work of  de Leeuw \textit{et al.}~\cite{leeuw}).
An alternative approach to deal with long-range systems
is proposed by Smith~\cite{Smith1994}. In his method, the Hamiltonian and
equations of motion are derived using constraints on the velocities of particles.
Instead, in the following we will stick to a standard model for the Hamiltonian
and will consider ways to improve the convergence in the potential energy.

For a good performance in simulations of large $N$-particle systems, a
number of  modified summation methods has been developed. Historically, the
first efforts to enhance the Ewald method consisted in looking for
appropriate truncation schemes, but all of them were strongly dependent on
the system properties, in particular on the system size. Tabulations of precalculated
terms in both real space and
momentum space sums~\cite{Sangester1976}, as well as polynomial
approximations of the involved
functions~\cite{Lage1947,Brush1966,Hansen1973}, were also proposed to look
for a balance between calculation time and truncation errors. Nevertheless,
these approximate methods suffer from error accumulation in simulations of
large systems, and do not allow for reducing the overall $\mathcal{O}(N^2)$
complexity of the original Ewald summation. The work of Perram \textit{et
al.}~\cite{Perram1988} was the first to give a way to optimise the
splitting of the interparticle potential between the long-range and
short-range parts to yield  a total complexity of $\mathcal{O}(N^{3/2})$.
A special modification of the Ewald method called Wolf
summation~\cite{wolf1999, wolf}, based on a damping of the
Fourier-transformed part of the sum, was posteriorly developed in order to
render the original Ewald summation more efficient for non-periodic systems
and large model sizes.

Another way for improving the Ewald method is to perform fast Fourier
transform (FFT) of a reciprocal space sum on a mesh. The oldest algorithm
of this kind is the so-called Particle-Particle Particle-Mesh (P$^3$M)
method, invented by Hockney and Eastwood in the late
80's~\cite{Hockney1988}. The P$^3$M technique is based on a distribution of
the charge density on a grid using a certain smooth assignment function and
then the discrete Poisson equation is solved using FFT. This algorithm
appeared to be less complex to yield $\mathcal{O}(N \ln N)$ with an
appropriate choice of the free parameters.
The P$^3$M algorithm was recently improved by Ballenegger
\textit{et al.}~\cite{Ballenegger2008} for calculation of energies,
bringing, as claimed, the maximal precision in the energy by an optimisation
of the ``influence'' function (a substitution of the potential in the
Fourier-transformed Poisson's equation).
For a comprehensive introduction
to Ewald- and mesh-based techniques  we recommend  to refer to the cited
work of Ballenegger and coauthors  where special attention is paid to
the estimation of both sum truncation-imposed and grid-imposed  errors.
The extension of this method,
called Particle Mesh Ewald~\cite{Darden1993} (PME), makes use of the
analytical form of the sum in the reciprocal space and evaluates potentials
via FFT instead of interpolating them as P$^3$M does. Although PME is
slightly more complex then the P$^3$M algorithm, it is still $\mathcal{O}(N
\ln N)$ and allows to reduce significantly the memory expenses. Later
Particle Mesh Ewald method was reformulated by Essmann \textit{et al.}~\cite{Essmann1995},
making use of cardinal {\it B}-splines to interpolate
structure factors. This approach, called Smooth Particle-Mesh Ewald (SPME)
substantially improved the accuracy of PME with a comparable computational
cost, as it still scales as $\mathcal{O}(N \ln N)$. SPME is also claimed to be
applicable to potentials of the polytropic form $1/|\bs{r}|^k$. In general,
the conventional FFT-based approaches suffer from the severe fallback of
requiring equidistant particle positions. The invention of the variant of
Fourier transform for nonequispaced nodes (NFFT) opened a path to overcome
this shortcoming, while keeping the introduced errors below the specified
target levels. The nonequispaced fast Fourier transform is currently
considered as a promising means to improve the Ewald summation performance,
with open code implementations available~\cite{nfftcode}. The early
variants of the NFFT algorithms are reviewed in the work of
A.~F.~Ware~\cite{Ware1998};  a general approach to the fast summation
methods based on NFFT can be found in the article of
G.~Steidl~\cite{steidl1998}.

The most recent family of algorithms based on the Ewald approach are
 the tree-based algorithms, with the fast multipole method (FMM) being the
 most known and widely used among them.  The algorithm, developed
 primarily by L.~Greengard and V.~Rokhlin~\cite{Greengard1987},
 is based on the idea of
keeping the direct summation of potentials or forces for the nearby atoms
and approximating the interactions of the distant atoms by their multipole
expansions. FMM offers the asymptotically fastest performance among the
Ewald-related algorithms, being linear in $N$ in most cases and not worse
than $\mathcal{O}(N \ln N)$ with explicitly controlled accuracy. The FMM
technique is naturally applicable to inhomogeneous and non-periodic
systems, being also easy to parallelise since it is an entirely real-space
summation.
Since then the algorithm was significantly improved in efficiency,
mostly by introducing new diagonal forms of translation operators~\cite{Greengard1997}.
However, FMM has an intrinsic shortcoming, when applied to molecular dynamics
calculations, as the energy conservation it brings is poor; the method
\textit{per se} is also rather cumbersome in implementation.
Another group of methods, based on the multigrid methods of solving
elliptic (in this particular case -- Poisson's) equations~\cite{Brandt1998},
was developed a decade ago~\cite{Sagui2001}. These methods allow to preserve
the scaling $\mathcal{O}(N)$ and parallelisation advantages of
tree-based methods, as well as the applicability in simulations without PBC,
being on the other hand satisfactorily energy-conserving and additionally
accelerated on all length scales.

  A detailed comparison of the optimised $\mathcal{O}(N^{3/2})$
pure Ewald technique, FFT-based summations, and multipole-based methods was
made by H.~G.~Petersen~\cite{Petersen1995} for systems with approximately
uniform charge distributions, taking into account a possible parallel
implementation. According to Petersen, the method of choice with a number
of particles below 10$^4$ is the conventional Ewald summation, PME is
preferable in the range $N\sim 10^4-10^5$, and the fast multipole method
should overperform them with $N >10^5$.
A more recent and ample review of FMM, P$^3$M and pure Ewald methods by
Pollock and Glosli~\cite{Pollock1996a}, based partially on
their own calculations, implies that P$^3$M is faster than the
Ewald summation already for 500 particles, although it is stressed that
the other factors as the ease of the coding, the system geometry, as well as
the code optimisation can change the choice. We would also suggest a thorough
survey of different Ewald summation techniques given in the work of Toukmaji
and Board~\cite{Toukmaji1996}.

An approach, alternative to using cubic periodic boundary conditions in a
calculation of long-range interactions, called Isotropic Periodic Sum
(IPS), was recently proposed by Wu and Brooks ~\cite{Wu2005}. The main goal
of their approach is to deal with long-range interactions, avoiding
artificial correlations and anisotropy bias
induced by a PBC-based summation in a cubic box.
In this technique, only the
interactions of a particle A with the others within a certain radius
$\bs{r}_c$ are taken into account (as in a plain cut-off scheme), and this
spherical simulation zone is repeated in an infinite number of shifts by
vectors $\bs{r}_{\rm sh}$, such that $|\bs{r}_{\rm sh}|=2N|\bs{r}_c|$.
Therefore, the particle A interacts not only with B (within the sphere
radius), but also with all the images of B, occupying homogeneously the
shells of radii $2N|\bs{r}_c|$, centered in B. The subsequent integration
and summation over the shells allows to obtain explicit expressions of
forces and energies for a number of interactions of most physical interest,
like electrostatic, Lennard--Jones and exponential potentials. The method
is known to yield a performance close to the one shown by the Ewald
summation, but without imposing unwanted symmetry effects.

Since its proposal, the Ewald method has been applied to a large number of
physical problems, although mostly to systems with the Coulomb $1/|\bs{r}|$
interaction potential. In a recent work by R.~E.~Johnson and
S.~Ranganathan~\cite{johnson}, a generalised approach to Ewald summation is
stated to obtain potential energy and forces for systems with a power-law,
Yukawa potential and electronic bilayer systems.
The Ewald method  for
two-dimensional systems with electrostatic interactions was
developed by Parry~\cite{Parry1975}, but his
technique appeared to be computationally inefficient.
Spohr \textit{et al}.~\cite{Spohr1997}  studied a slab geometry  by treating the
simulation cell as a fully three-dimensional one with the conventional Ewald
summation. Later on, a significant advance was made by Yeh and
Berkowitz~\cite{Yeh1999}, as
the authors managed to obtain the explicit correction term for the
rigorous three-dimensional Ewald summation, that brings the results for
a slab system in a satisfactory agreement with the 2D summation.
The 2D Ewald technique was also applied by Wen Yang \textit{et
al.}~\cite{wenyang}
to calculate the energy of Coulomb particles in a slab system with a
uniformly charged
surface. Recent applications to dipolar bosons in a 2D geometry
have been made by  C.~Mora \textit{et al.}~\cite{mora} and Xin Lu
\textit{et al.}~\cite{Xin2008}. On the other hand, the explicit forms of
the Ewald sums for Yukawa interactions have been also reported: in 3D
geometry, with partial periodical boundary conditions~\cite{mazars1}, and in 2D
geometry~\cite{mazars2}.
The Ewald method can also be useful even applied to fast decaying power-law
potentials. For instance, Shirts \textit{et al.} in their recent work~\cite{Shirts2007}
argue the need for taking into account the effects of cutoffs
in molecular dispersion interactions due to a Lennard-Jones potential,
especially in non-isotropic and inhomogeneous media. The authors developed
two formalisms for the estimation of these cutoff errors in binding free
energy of macromolecular systems, which can in principle be extended to the
other observables. However, it is claimed that the adequate implementation
of the Ewald summation for this kind of systems may render their
corrections unnecessary by mostly eliminating the cutoff-dependent
behaviour.

In the present work, we report explicit expressions of the Ewald sums for
the general case of particles interacting via a $1/|\bs{r}|^k$ polytropic
potential and in 3D, 2D, and 1D geometries.  The closed derivation of these
sums is given, with special attention being paid to conditionally convergent
potentials.  One of the difficulties of the derivation is that different
terms have to be considered in the cases of short-range, long-range or
``marginal'' potentials. In the case of a short-range interaction, the
original slowly convergent sum is represented as a linear combination of
two rapidly convergent ones. For a long-range interaction, the condition of
charge neutrality in the simulation cell is shown to be necessary to make the energy absolutely convergent within the considered
scheme. The introduction of a uniform neutralizing charged background
(\textit{jellium}), as a particular case of a charge-neutral system, is
also discussed. The explicit forms of the Ewald sums are reported for a
jellium system and for an arbitrary polytropic potential. We explicitly
calculate the expressions for physically relevant interactions as Coulomb,
dipole-dipole, and Lennard-Jones potentials. Finally, we have extended the
Ewald sums to the case of   a noncubic simulation cell, that could be
useful in simulations of hexagonal  closed packed (hcp) and two-dimensional
triangular solids. In addition, the general derivation path given in this
work may be used to obtain the forms of Ewald sums for other interaction
potentials.

The computational efficiency is another important issue of the practical
implementation of the method. In fact, one needs to choose correctly a free
parameter, appearing in the integral representation of the sums, and to
decide which number of terms should be kept in spatial and momentum sums in
order to reach the required accuracy. The choice of these three parameters
affects the difference between the calculated result and the exact one as
well as  the calculation complexity. Therefore, a certain optimisation of
the parameters is always required. In the present work, this optimisation
process is formalised and it is shown that following the described
procedure the overall computation time is significantly reduced. The
accuracy of the result is shown to be kept under control, with the only
cost of a preliminary benchmark
calculation.

The rest of the article is organised as follows. In
Section~\ref{sec:solforarb}, we formulate the problem, develop the general
Ewald approach and report explicit expressions for the Ewald sums for a
polytropic potential in a three-dimensional cubic simulation cell.
Sections~\ref{sec:ewald2d} and~\ref{sec:ewald1d} contain derivations of the
Ewald sums in two-dimensional and one-dimensional geometries, respectively.
In Section~\ref{sec:ewaldbox}, the case of a simulation cell with different
side lengths is considered for three- and two-dimensional systems. The
final general expressions and their particularization to the  most
physically relevant  cases are presented in Section~\ref{sec:results}. The
practical algorithm for the parameter optimisation and an actual
application of the Ewald method is discussed in
Section~\ref{sec:practical}. Summary and conclusions  are drawn in
Section~\ref{sec:conclusions}.

\section{Ewald sum for an arbitrary polytropic potential $1/|\bs{r}|^k$ in 3D geometry}
\label{sec:solforarb}

\subsection {Basic assumptions and initial sums}
We consider a system of $N$ particles inside a cubic simulation cell of
size $L$ with periodic boundary conditions. Thus,
each particle with coordinates $\bs{r}$ in the initial cell has an
infinite number of images $\bs{r}+\bs{n}L$ in the adjacent cells. The
total potential energy is estimated by
\begin{equation}
\Pi=\frac{1}{2}{\sum_{\bs{n} \in \mz^3}} ^{\prime}
\left[\sum_{i=1}^{N}\sum_{j=1}^{N}\phi(\bs{r}_{ij}+\bs{n}L)\right]
\label{initialhamiltonian1}
\end{equation}
where $\phi(\bs{r})$ is the interparticle potential, $\bs{r}_{ij} \equiv \bs{r}_i -
\bs{r}_j$, and the prime in the first sum means that the summation over
an integer vector $\bs{n}$ must be done omitting the term $\bs{n}=\bs{0}$ when $i=j$.

\subsection{Analytic development}
In many physical situations, the interaction potential between two particles
$i$ and $j$ has  the power-law form $q_i q_j/|\bs{r}|^k$ with positive $k$
and $q_i$, $q_j$ being the generalised charges of the particles. This sort of
interaction is generally referred to as \textit{polytropic} potential.

First, let us consider the case of short-range potentials, $k \leq 3$. As we
will see later, the potentials corresponding to $k>3$ give a similar
result. For $k \leq 3$, the right-hand part of
Eq.~(\ref{initialhamiltonian1}) diverges and it can be made
convergent only if the restriction of charge neutrality is required, i.e., when
$\sum_{i=1}^{N}q_i=0$.
It has also been shown~\cite{Fraser1996} that for a pure electrostatic
interaction  the total energy (\ref{initialhamiltonian1}) can be
conditionally convergent even in  a neutral simulation cell because of a
higher multipole contribution. The energy and forces  are therefore
dependent on the order of summation, which can also be implicitly set by a
choice  of a convergence factor. The ambiguity usually appears in a form of
a constant or a position-dependent term, vanishing in the limit
$L\rightarrow \infty$. Hence, the preference in one or another factor should be
dictated either by physical properties of a particular system or by arguments
regarding rates of convergence to the thermodynamic limit. For a general
discussion on the convergence issues appearing in periodic boundary
conditions, see Ref.~\cite{Makov1995}.
The main idea of the Ewald summation technique in the approach proposed by
de Leeuw, Perram, and Smith~\cite{leeuw} is to multiply each component of
the sum by the dimensionless factor $e^{-s n^2}$, with $s>0$ being a
dimensionless regularizing parameter, making the sum absolutely convergent.
Then, the limit $s \rightarrow 0$ is taken, so that the singularity in the
initial sum~(\ref{initialhamiltonian1}) can be explicitly separated into a
term depending only on $s$, that finally can be  cancelled due to the
charge neutrality condition. We take a similar multiplier
$c(\bs{n},\bs{r},s)=e^{-s |\bs{n}+\bs{r}|^2}$ yielding the same rate of
convergence (since $0\leq r\leq 1$ in units of $L$). As the sum, multiplied
by $c$, is invariant to an arbitrary substitution
$\bs{r}\rightarrow\bs{n}+\bs{r}$, the chosen convergence factor allows to
preserve the periodicity of the potential in order to avoid any possible
artefacts in the final results.

For the sake of clearness of the derivation, it is convenient to use
reduced length units, that is to use the size of the box $L$ as unity of
length and substitute $r_{ij}$ by $r_{ij}L$. From now on, and for simplicity,
we use the notation $\bs{r}$ for
$\bs{r}_{ij}$ and, in case of possible ambiguity, we will stick to the
standard notation $\bs{r}_{ij}$.
Also, we rewrite the potential energy by splitting the total sum
(\ref{initialhamiltonian1}) into two terms:
$I_{01}$ (the sum of the interactions between a particle with
all the \textit{other} particles in the box), and $I_{00}$ (the
sum of the interaction of a particle with \textit{its own} images, comprised of the components $i=j$ in Eq. (\ref{initialhamiltonian1})).
Explicitly,
\begin{equation}
\Pi = \frac{1}{L^k}(I_{01} + I_{00}) \ ,
\label{ep1}
\end{equation}
with
\begin{eqnarray}
I_{01} & =& \sum_{\bs{n} \in \mz^3} \left[\sum_{1\leq i<j\leq N}
\frac{q_i q_j e^{-s |\bs{r}_{ij}+\bs{n}|^2}}{|\bs{r}_{ij}+\bs{n}|^k}\right] \label{i01init}\\
I_{00} &= & \frac{1}{2}\sum_{\bs{n} \in \mz^3\backslash \bs{0}}
\frac{e^{-s n^2}}{n^k}\sum_{i=1}^{N}q_i^2 \  ,
\label{i00init}
\end{eqnarray}
where the shorthand notation $n=|\bs{n}|$ is used.

First, let us focus on the $I_{01}$ term, which we rewrite as
\begin{eqnarray}
I_{01}=\sum_{1\leq i<j\leq N}q_i q_j \psi(\bs{r},s) \ ,
\label{tildei10polytropic}
\end{eqnarray}
where we have defined the ``screened'' interaction potential
$\psi(\bs{r},s)=\sum_{\bs{n}}e^{-s |\bs{r}+\bs{n}|^2}/|\bs{r}+\bs{n}|^k$,
extended from a single cell to the whole coordinate space.
Since the total potential energy consists of a sum of pair interaction
components, we may consider a single pair without any loss of generality.

Let us apply the equation
\begin{equation}
x^{-2s}=\frac{1}{\Gamma(s)}\int_0^{\infty}t^{s-1}e^{-tx^2}\di t \ ,
\label{repx1}
\end{equation}
representing the definition of the gamma-function, to the polytropic potential $|\bs{r}+\bs{n}|^{-k}$.
Then the function $\psi$ may be represented in an integral form,
\begin{equation}
\psi(\bs{r},s)=\frac{1}{\Gamma(k/2)}\int_0^{\infty}t^{\frac{k}{2}-1}
\sum_{\bs{n}}e^{-t|\bs{r}+\bs{n}|^2}e^{-s |\bs{r}+\bs{n}|^2}\di t \ .
\label{boro1}
\end{equation}

We expect that the integral (\ref{boro1}) contains a singularity
that will be located in the vicinity of zero. Therefore,
we split this integral into two domains $[0,\alpha^2]$ and
$[\alpha^2,\infty)$,
the corresponding integrals being denoted as $\psi_{{\rm fin}}$ and
$\psi_{{\rm inf}}$, where $\alpha$ is some arbitrary positive
constant,
\begin{equation}
   \psi(\bs{r},s)  = \psi_{{\rm fin}}(\bs{r},s) + \psi_{{\rm inf}}(\bs{r},s) \ .
\label{boro2}
\end{equation}
In the following, we analyze the two terms of the previous sum
(\ref{boro2}).

\begin{enumerate}
\item
The explicit analytical form of the term $\psi_{{\rm inf}}(\bs{r},s)$ can
be found
\begin{equation}
 \psi_{{\rm inf}}(\bs{r},s) =\frac{1}{\Gamma (\frac{k}{2})}
 \sum_{\bs{n}}\int_{\alpha^2}^{\infty}t^{\frac{k}{2}-1}e^{-t|\bs{r}+\bs{n}|^2-s|\bs{r}+\bs{n}|^2}\di t=
\sum_{\bs{n}}\frac{e^{-s |\bs{r}+\bs{n}|^2 }}{|\bs{r}+\bs{n}|^k}
\frac{\Gamma (\frac{k}{2},\alpha ^2|\bs{r}+\bs{n}|^2 )}{\Gamma
(\frac{k}{2})}  \ ,
\end{equation}
where $\Gamma(a,z)$ is the incomplete gamma function. From the large
distance asymptotic expansion of this function, one obtains that the above lattice
sum is absolutely and uniformly convergent if $s \geq 0$ and $\alpha>0$.
Therefore, one may simply take the limit of vanishing
screening $s \rightarrow 0$,
\begin{equation}
\psi_{{\rm inf}}(\bs{r},s) \lims \frac{1}{\Gamma (\frac{k}{2})}\sum_{\bs{n}}
\frac{\Gamma (\frac{k}{2},\alpha ^2|\bs{r}+\bs{n}|^2 )}{|\bs{r}+\bs{n}|^k}
\ .
\label{i01infpolytropic}
\end{equation}

\item
The calculation of  $\psi_{{\rm fin}}(\bs{r},s)$ is done
by making a separate analysis of the $\bs{n}=\bs{0}$ case,
\begin{equation}
\psi_{{\rm fin}}(\bs{r},s) = \psi_{{\rm fin}}^{\bs{n}\neq \bs{0}}(\bs{r},s)
+\psi_{{\rm fin}}^{\bs{n}=\bs{0}}(\bs{r},s) \ . \label{psifinsplit}
\end{equation}
Explicitly,
\begin{eqnarray}
 \psi_{{\rm fin}}^{\bs{n}\neq \bs{0}}(\bs{r},s)  & = &\frac{\pi^{\frac{3}{2}}}{\Gamma(\frac{k}{2})}
\sum_{\bs{n}\neq \bs{0}} \int_0^{\alpha^2}\frac{t^{\frac{k}{2}-1}}{(t+s)^{\frac{3}{2}}}
\exp\left[\frac{-\pi^2n^2}{t+s}+2\pi \ii
\bs{n}\bs{r}\right]\di t \label{psifinnneq0}\\
\psi_{{\rm fin}}^{\bs{n}=\bs{0}}(\bs{r},s)& = &\frac{\pi^{\frac{3}{2}}}{\Gamma(\frac{k}{2})}
\int_0^{\alpha^2}\frac{t^{\frac{k}{2}-1}}{(t+s)^{\frac{3}{2}}}
\di t \ ,
\label{i11tildawodiv2}
\end{eqnarray}
where we have used the Jacobi transformation~\cite{Jacobibook,Whittaker1999}
\begin{equation}
\sum_{\bs{n}} e^{-s|\bs{n}+\bs{r}|^2}=\left(\frac{\pi}{s}\right)^{3/2}
\sum_{\bs{n}}\exp[-\pi^2 n^2/s +2\pi \ii \bs{n}\bs{r}]\;\;{\rm for} \;\;
\bs{n}\in \mz^3  \ ,
\label{3djacobi}
\end{equation}
applied to
\begin{equation}
\exp[-s |\bs{n}+\bs{r}|^2-t|\bs{n}+\bs{r}|^2]=
\exp[-(s+t)|\bs{n}+\bs{r}|^2].
\label{transf1}
\end{equation}

We evaluate the integral $\psi_{{\rm fin}}^{\bs{n}\neq \bs{0}}(\bs{r},s)$
by the following analysis. Consider separately the following factor of the integrated expression from (\ref{psifinnneq0})
\begin{equation}
M=\frac{\exp\left[-\frac{\pi^2n^2}{t+s}\right]}{(t+s)^{\frac{3}{2}}} .
\end{equation}
It is clearly continuous and bounded on $(0,\:+\infty)$ as a function of $(t+s)$,
also notice that $t^{k/2-1}$ is absolutely integrable on $(0,\:\alpha^2)$ for $k>0$.
In accordance with the standard convergence test for improper integrals,
the integral $\psi_{{\rm fin}}^{\bs{n}\neq \bs{0}}(\bs{r},s)$ converges absolutely and uniformly
with $s$ being considered as a parameter. Then, the limit $s\rightarrow 0$ may be carried out
and the integral becomes
\begin{eqnarray}
\psi_{{\rm fin}}^{\bs{n}\neq \bs{0}}(\bs{r},s)=&&\frac{\pi^{\frac{3}{2}}}{\Gamma(\frac{k}{2})}\sum_{\bs{n}\neq \bs{0}}e^{2\pi \ii \bs{n}\bs{r}} \int_{0}^{\alpha^2}t^{\frac{k-5}{2}} \exp\left[-\frac{\pi^2n^2}{t}\right] \di t \label{i11gc1} \\
=&&\sum_{\bs{n}\neq \bs{0}}\frac{\pi^{\frac{3}{2}}\cos(2\pi \bs{n}\bs{r})}{\Gamma(\frac{k}{2})}
\alpha^{k-3} E_{\frac{k-1}{2}}\left(\frac{\pi ^2n^2}{\alpha ^2}\right) \ . \label{i01nnot0fin}
\end{eqnarray}
The function $E_{n}(z)$ is the exponential integral function, and we have cancelled the imaginary
part of the sum (\ref{i11gc1}) by grouping the pairs with $\bs{n}$ and $-\bs{n}$.

Now, we analyze the second term of $\psi_{{\rm fin}}(\bs{r},s) $,
\begin{equation}
 \psi_{{\rm fin}}^{\bs{n} = \bs{0}}(\bs{r},s)  = \frac{\pi^{\frac{3}{2}}}
 {\Gamma(\frac{k}{2})} \int_0^{\alpha^2}\frac{t^{\frac{k}{2}-1}}{(t+s)^{\frac{3}{2}}}
 \di t  \ .
\end{equation}
In terms of a new variable $v=s/(t+s)$,
\begin{equation}
 \psi_{\rm fin}^{\bs{n} = \bs{0}}(\bs{r},s)  = \frac{\pi^{\frac{3}{2}}}
 {\Gamma(\frac{k}{2})} \int_{s/(\alpha^2+s)}^1\frac{(1-v)^{\frac{k}{2}-1}}
 {v^{\frac{k-1}{2}}} s^{(k-3)/2}\di v \ .
 \label{i11n0withexp}
\end{equation}

The integration of $\psi_{{\rm fin}}^{\bs{n} = \bs{0}}(\bs{r},s)$
for a $1/|r|^k$ interaction has to be carefully analyzed as a function of
$k$: $1\leq k<3$, long-range potential; $k=3$, marginal case; and $k>3$,
short-range potential.
\begin{enumerate}
%
%
\item
Suppose $1\leq k<3$. The resulting integral,
\begin{equation}
\psi_{{\rm fin}}^{\bs{n} = \bs{0}}(\bs{r},s) =
 \psi_{\rm fin}^{\bs{n} = \bs{0}}(\bs{r},s)  = \frac{\pi^{\frac{3}{2}}}
 {\Gamma(\frac{k}{2})} \int_{s/(\alpha^2+s)}^1\frac{(1-v)^{\frac{k}{2}-1}}
 {v^{\frac{k-1}{2}}} s^{(k-3)/2}\di v
\label{i11gc2}
\end{equation}
may be given explicitly in terms of incomplete beta- and incomplete
gamma-functions. Expanding the resulting function for small $s$,
\begin{equation}
\psi_{{\rm fin}}^{\bs{n} = \bs{0}}(\bs{r},s) =s^{\frac{k-3}{2}} 2 \pi
\Gamma\left[\frac{3-k}{2}\right]+\frac{2 \pi ^{\frac{3}{2}} \alpha ^{k-3}}{(k-3)
\Gamma\left[\frac{k}{2}\right]} + \mathcal{O}(s)
\label{i01klt3}
\end{equation}
It is easily seen, that the only divergent term in the
expansion~(\ref{i01klt3}) is the first one, which we define as
\begin{equation}
S(s)=s^{\frac{k-3}{2}} 2 \pi  \Gamma\left[\frac{3-k}{2}\right]\ .
\label{Ssklt3}
\end{equation}

We remind that the choice of a convergence factor (that explicitly affects the
summation order) may in principle lead to additional contributions in the total
energy if the convergence of the sum is conditional (like for a
charge-neutral cell of Coulomb particles with non-zero total dipole
moment). In the original derivation of de Leeuw et al.~\cite{leeuw}, the
factor $\exp(-sn^2)$ results in an additional dipole-like
component in $\psi_{\rm fin}^{\bs{n} = \bs{0}}$, which breaks the
periodicity of the potential and therefore complicates its use
in simulations with periodic boundary conditions. Moreover, this procedure~\cite{leeuw}
yields a nonvanishing dipole term exclusively for $k=1$ in 3D geometry,
with the rest of the sums remaining unchanged. From our point of view, this
discontinuity points out to an nonphysical character of the dipole term
appearing in the case of the Coulomb potential. Nevertheless, in a number
of studies~\cite{Fraser1996, Makov1995} it is considered as a first order
correction when the convergence to the thermodynamic limit is analyzed. The
mere fact that the results for the two different convergence multipliers
coincide when $k>1$ is a consequence of the absolute convergence of the
higher multipole contributions in this case.

\item Suppose $k=3$. In this marginal case, the expression~(\ref{i11gc2})
may be integrated directly to yield the following logarithmic dependence
\begin{equation}
\psi_{{\rm fin}}^{\bs{n} = \bs{0}}(\bs{r},s)  = \frac{\pi^{\frac{3}{2}}}{\Gamma(\frac{3}{2})}\left(\frac{-2\alpha s-2\alpha^3}{(\alpha^2 +s)^{\frac{3}{2}}}+\ln(s+2\alpha^2+2\alpha\sqrt{s+\alpha^2})-\ln s\right)
\end{equation}
that close to $s=0$ expands as
\begin{equation}
 \psi_{{\rm fin}}^{\bs{n} = \bs{0}}(\bs{r},s) =-2\pi\ln\,s-4\pi+4\pi \ln (2\alpha) + \mathcal{O}(s \ln s)\label{i01keq3}
\end{equation}
with the diverging term
\begin{equation}
S(s)=-2\pi\ln\,s\ . \label{Ssk3}
\end{equation}
%
%
\item
Consider the remaining option $k>3$. In this case,
$(1-v)^{\frac{k}{2}-1}$ is bounded from above  and $(k-1)/2>1$. It means
that the integral converges absolutely and the only finite contribution to the
integral comes from the first (constant) term of the integral expansion
for small $s$,
\begin{equation}
\psi_{{\rm fin}}^{\bs{n} = \bs{0}}(\bs{r},s) = \frac{2 \pi ^{\frac{3}{2}}
\alpha ^{k-3}}{(k-3) \Gamma\left[\frac{k}{2}\right]}
\label{i01kgt3}
\end{equation}

\end{enumerate}
\end{enumerate}

The second term of the total potential energy, $I_{00}$ (\ref{ep1}) can be
derived in a similar form to the first one.
The procedure to find the form of $\psi(\bs{r},s)$ is repeated here with
$\bs{r}_{ij}=0$, hence the results are obtained straightforwardly via
(\ref{i01infpolytropic}), (\ref{i01nnot0fin}), (\ref{i01keq3}) and
(\ref{i01kgt3}),
\begin{eqnarray}
I_{00}=&&\sum_{i=1}^{N}q_i^2\left[\frac{1}{\Gamma (\frac{k}{2})}\sum_{\bs{n}}\frac{\Gamma
(\frac{k}{2},\alpha ^2 n^2 )}{n^k}+\sum_{\bs{n}\neq \bs{0}}\frac{\pi^{\frac{3}{2}}}
{\Gamma(\frac{k}{2})}
\alpha^{k-3} E_{\frac{k-1}{2}}\left(\frac{\pi ^2 n^2}{\alpha ^2}\right)\right. \nonumber \\
&&\left.-\frac{\alpha^k}{\Gamma(\frac{k}{2}+1)}+\psi_{{\rm fin}}^{\bs{n} = \bs{0}}(\bs{r},s)
\right] \ ,
\label{i00poly}
\end{eqnarray}
with the term $\psi_{{\rm fin}}^{\bs{n} = \bs{0}}(\bs{r},s)$ depending on the
potential parameter $k$ via (\ref{i01klt3}), (\ref{i01keq3}) or
(\ref{i01kgt3}).

Putting all together, the potential energy can be written in a more compact form as,
\begin{eqnarray}
\Pi=&&\frac{1}{L^k}(I_{01}+I_{00})\nonumber\\
    =&&\frac{1}{L^k}\sum_{i<j}q_i q_j \psi(r_{ij}/L)+\frac{1}{2L^k}
    \sum_{i=1}^{N} q_i^2 \, \xi+
\frac{1}{L^k}\sum_{i<j}q_i q_j S(s)+\frac{1}{2L^k}\sum_{i}q_i^2 S(s) \ ,
\end{eqnarray}
with the generalised potential,
\begin{equation}
\psi(\bs{r})= \sum_{\bs{n}}R(\bs{n},\bs{r})+\sum_{\bs{n}\neq
0}K(\bs{n},\bs{r})+C_1 \ .
\label{psipoly1}
\end{equation}
A constant shift in the definition of
$\psi$ is introduced to satisfy by the property $\int_{\rm cell}\psi \di \bs{r}=0$,
convenient for a proper treatment of  the background contributions (see
Appendix). The functions entering in Eq. (\ref{psipoly1}) are defined as
\begin{eqnarray}
R(\bs{n},\bs{r})&=&  \frac{\Gamma (\frac{k}{2},\alpha ^2 |\bs{r} + \bs{n}|^2 )}
{\Gamma (\frac{k}{2})|\bs{r} + \bs{n}|^k}  \label{defr}\\
K(\bs{n},\bs{r})&=&\kappa(\bs{n}) \, \cos(2\pi \bs{n}\bs{r}) \ , \label{defk}\\
\end{eqnarray}
with
\begin{equation}
\kappa(\bs{n})=\frac{\pi^{\frac{3}{2}}\alpha^{k-3}}{\Gamma(\frac{k}{2})}
 E_{\frac{k-1}{2}}\left(\frac{\pi ^2n^2}{\alpha ^2}\right) \ .
 \label{defkappa}
\end{equation}
The explicit form of the function $S(s)$ depends on the $k$ value,
\begin{equation}
S(s)=\left\{
\begin{array}{ll}
s^{\frac{k-3}{2}} 2 \pi  \Gamma\left[\frac{3-k}{2}\right]\, &\textrm{if $k \leq 3$ (singular term)}
\label{singterm}\\
-2\pi \ln s\, &\textrm{if $k=3$ (singular term)} \\
0\, &\textrm{if $k>3$}\end{array}\right. \ ,
\end{equation}
and the term $\xi$ depends only on the choice of $\alpha$,
\begin{equation}
\xi = \sum_{\bs{n}\neq
0}(\rho(\bs{n})+\kappa(\bs{n}))+C_1+C_2 \ ,
\label{xipoly1}
\end{equation}
with
\begin{equation}
\rho(\bs{n})=\frac{\Gamma (\frac{k}{2},\alpha ^2n^2 )}{\Gamma
(\frac{k}{2})n^k} \ ,
\label{defrho}
\end{equation}
and $\kappa(\bs{n})$ defined in Eq. (\ref{defkappa}). The constants $C_1$
and $C_2$ are explicitly,
\begin{eqnarray}
C_{1}&=&\left\{
\begin{array}{ll}
\frac{2 \pi ^{\frac{3}{2}} \alpha ^{k-3}}{(k-3) \Gamma\left[\frac{k}{2}\right]} & \textrm{if $k \neq 3$}\\
-4\pi+4\pi\ln(2\alpha) & \textrm{if $k=3$}\\
\end{array}\right. \label{c1} \\
C_{2}&=&-\frac{\alpha^k}{\Gamma(\frac{k}{2}+1)}\label{c2}\\
\end{eqnarray}

\subsection{Removing singularities for $k\leq 3$}
\label{sec:remsing}

The diverging part $\Pi_s$ (containing a singularity) of the total potential
energy equals to
\begin{equation}
\Pi_s=\frac{1}{L^k}\sum_{i<j}q_i q_j S(s)+\frac{1}{2L^k}\sum_{i}q_i^2 S(s)
=\frac{1}{2L}\left(\sum_{i}q_i\right)^2 S(s)
\label{epotsingular}
\end{equation}
and vanishes, if the charge neutrality condition $\sum_{i}q_i=0$ is taken.

Consider now a charge-neutral system with a neutralizing background consisting
of a large number of identical uniformly distributed
particles of the opposite charge (the ``jellium'' model). We denote the numbers of negatively
charged particles $q_{-}$ and positively charged (background) particles $q_{+}$
as $N_{-}$ and $N_{+}$, respectively. By imposing charge neutrality,
$q_{+}=-[N_{-}/N_{+}]q_{-}$, with $N$ the total number of particles, $N=N_{-}+N_{+}$.

The potential energy for the jellium model can be written as
\begin{equation}
\Pi = \frac{1}{L^k}\sum_{i<j}q_i q_j
\psi(r_{ij}/L)+\frac{N_{-}q_{-}^2+N_{+}q_{+}^2}{2L^k}\xi \ .
\label{ep2}
\end{equation}

The second term in Eq. (\ref{ep2}) has a component proportional to $N_{+}q_{+}^2$.
Note that the negative charges $q_{-}$ and their number $N_{-}$ is defined by the problem and therefore fixed. Hence, in the limit $N_{+} \rightarrow \infty$, this term cancels
$N_{+}q_{+}^2 = (N_{-}^2 q_{-}^2)/N_{+} \rightarrow 0 $, and therefore this background contribution
may be eliminated to yield
\begin{equation}
\frac{N_{-}q_{-}^2+N_{+}q_{+}^2}{2L^k}\xi = \frac{N_{-}q_{-}^2}{2L^k} \xi \ .
\end{equation}
Concerning the first term of Eq.~(\ref{ep2}), let us split it into three
pieces,
\begin{equation}
\frac{1}{L} \sum_{1\leq i<j\leq N}q_i q_j \psi(\bs{r}) =
\frac{1}{L}(S_{--}+2 S_{-+}+S_{++}) \ ,
\label{psibythrees}
\end{equation}
where the first sum corresponds to the interaction between the negative charges
\begin{equation}
S_{--}=\sum_{1\leq i<j\leq N_{-}}q_i q_j \psi(\bs{r}) \ ,
\label{psisum1}
\end{equation}
the second sum is the interaction of the negatively charged particles with the positive charges of the background
\begin{equation}
S_{-+}=\sum_{i=1}^{N_{-}}\sum_{j=1+N_{-}}^{N_{+}+N_{-}}q_i q_j
\psi(\bs{r}) \ ,
\label{psisum2}
\end{equation}
and the third one is the interaction between the background charges
\begin{equation}
S_{++}=\sum_{1+N_{-}\leq i<j\leq N_{+}+N_{-}}q_i q_j \psi(\bs{r}) \ .
\label{psisum3}
\end{equation}
The last two terms $S_{-+}$ and $S_{++}$ are easily shown to be zero in the limit
$N_+\rightarrow \infty$ as a consequence of the zero value of the integral of
$\psi$ over the simulation cell (see Appendix).

With the above considerations we can finally write the expression for the
potential energy within the jellium model as
\begin{equation}
\Pi^{{\rm jel}} = \frac{q_{-}^2}{L^k}\sum_{i<j}\psi(r_{ij}/L)+\frac{Nq_{-}^2}{2L^k}\xi
\label{epgenfin0}
\end{equation}

In the more general case of different charges in a charge-neutral
simulation cell (with a long-range potential) or a system with an arbitrary
short-range potential the potential energy is given by
\begin{equation}
\Pi^{{\rm gen}} = \frac{1}{L^k}\sum_{i<j}q_i
q_j\psi(r_{ij}/L)+\frac{\sum_{i=1}^{N}q_i^2}{2L^k}\xi \ .
\label{epgenfin}
\end{equation}

A certain analytical conversion of the sum in the reciprocal space is
also possible in order to sum it up faster. Expanding the sum that defines
$K(\bs{n},\bs{r})$ (\ref{defk}), one can simplify it in the following way,
\begin{eqnarray}
\lefteqn{\sum_{i<j}q_i q_j \sum_{\bs{n}\neq 0}K(\bs{n},\bs{r})=\sum_{\bs{n}\neq 0}\kappa(\bs{n})
\sum_{i<j}q_i q_j \cos(2\pi \bs{n}\bs{r})} \nonumber\\
&&=\frac{1}{2}\sum_{\bs{n}\neq 0}\kappa(\bs{n})\sum_{i,j}q_i q_j\left[\cos(2\pi \bs{n}\bs{r}_i)
\cos(2\pi \bs{n}\bs{r}_j)+\sin(2\pi \bs{n}\bs{r}_i)\sin(2\pi
\bs{n}\bs{r}_j)\right] -\frac{1}{2}\sum_{i}q_i^2\sum_{\bs{n}\neq
0}\kappa(\bs{n}) \nonumber \\
&&=\frac{1}{2}\sum_{\bs{n}\neq 0}\kappa(\bs{n})
\left| \sum_{j}q_j\exp(2\pi \ii \bs{n}\bs{r}_j)\right|^2-\frac{1}{2}\sum_{i}q_i^2\sum_{\bs{n}\neq 0}\kappa(\bs{n})
\label{recipexpansion}
\end{eqnarray}
In this form, the sum over all pairs of particles in the reciprocal space
is represented as a single sum over particles and thus it scales as
$\mathcal{O}(N)$ instead of $\mathcal{O}(N^2)$.
Notice that the number of prefactors $\kappa(\bs{n})$ and exponents in the
sum depends on a chosen cutoff, which in general also might depend on $N$,
making the overall complexity of the $k$-space grow.
Na\"{\i}ve schemes with $\alpha$ and the cutoff not depending on $N$ do not
take into account the interplay between the $r$-space and $k$-space sum
complexities, thus leaving at least $\mathcal{O}(N^2)$ in one of them.
Nevertheless, as we show later, optimisation with $\alpha$ and cutoff
depending on $N$ gives a best total complexity of $\mathcal{O}(N^{3/2})$.
An alternative method to sum up the momentum space part is to use Fast
Fourier transform-based techniques (like PME), which is fast as
$\mathcal{O}(N \ln N)$.

The last term in Eq. (\ref{recipexpansion}) cancels  the $\kappa(\bs{n})$
component of $\xi$. Introduce the notation,
\begin{eqnarray}
\tilde{\psi}(\bs{r})&&=\sum_{\bs{n}}R(\bs{n},\bs{r})+C_1\label{tildepsi}\\
\tilde{\xi}&&=\sum_{\bs{n}\neq \bs{0}}\rho(\bs{n})+C_1+C_2\label{tildexi}\\
\tilde{S}_{\rm equal}(\bs{n})&&=q_{-}\sum_{j}\exp(2\pi \ii \bs{n}\bs{r}_j/L)\label{sq}\\
\tilde{S_q}(\bs{n})&&=\sum_{j}q_j\exp(2\pi \ii \bs{n}\bs{r}_j/L) \ ,
\label{se}
\end{eqnarray}
where $\tilde{S}_{\rm equal}$ is used when the system of equally charged particles $q_{-}$ is considered.
 Within this notation the potential energy may be rewritten in the following forms, which are more efficient for
numerical implementation,
\begin{eqnarray}
\Pi^{{\rm jel}} &&= \frac{q_{-}^2}{L^k}\sum_{i<j}\tilde{\psi}(r_{ij}/L)+
\frac{1}{2L^k}\sum_{\bs{n}\neq 0}\kappa(\bs{n})|\tilde{S}_{\rm equal}(\bs{n})|^2+\frac{Nq_{-}^2}{2L^k}\tilde{\xi}\label{epgenfin0eff}\\
\Pi^{{\rm gen}} &&= \frac{1}{L^k}\sum_{i<j}q_i q_j\tilde{\psi}(r_{ij}/L)+
\frac{1}{2L^k}\sum_{\bs{n}\neq
0}\kappa(\bs{n})|\tilde{S_q}(\bs{n})|^2+\frac{\sum_{i=1}^{N}q_i^2}{2L^k}\tilde{\xi} \ ,
\label{epgenfineff}
\end{eqnarray}
with $\bs{r}_i,\:r_{ij}$ in the original length units.

\subsection{Short-range potentials and the marginal case}
In case of a short-range interaction ($k>3$), the potential
energy does not diverge, which is clear from the form of the singular term
$S(s)$(see Eq. \ref{singterm}).
Hence, there is no need to add a neutralizing
background and, even more, the background must be necessarily excluded since it
leads to a divergence in the energy. This is easily seen by considering the
potential energy of the background separately
\begin{equation}
\Pi_{bg}=C \int_{0}^{\rm cell}\frac{\di \bs{r}}{|\bs{r}|^k} \ ,
\end{equation}
that contains a singularity in zero.
The expression for the potential energy is simply equal to
Eq.~(\ref{epgenfin}),
\begin{equation}
\Pi = \frac{1}{L^k}\sum_{i<j}q_i q_j
\psi(r_{ij}/L)+\frac{\sum_{i=1}^{N}q_i^2}{2L^k}\xi \ .
\label{ep22}
\end{equation}

When $k=3$ (marginal case), both ultraviolet and infrared divergences arise
in zero for the
background as well as in the vicinity of infinity (the logarithmic
divergence in the energy of negative charges). The only coherent model
here is a plain ``quasi-neutral'' gas consisting of a mixture of a finite number of
charges per box with the constraint $\sum q_i = 0$, i.e.,  with the positive
background excluded.

%
%
%
\section{Ewald method for two-dimensional systems}
\label{sec:ewald2d}

\subsection{General notes for lower dimensions}
The Ewald sums can be extended to two-dimensional (2D) systems interacting
through polytropic potentials.
The difference with the 3D case comes from a different form of the Jacobi imaginary
transformation for the Jacobi $\theta$-functions [its 3D form is given in
Eq.~(\ref{3djacobi})].

The ``third'' Jacobi $\theta$-function $\theta_3(z,\tau)$ is defined as
\begin{equation}
\theta_3(z|\tau)=\sum_{n=-\infty}^{+\infty}e^{\ii\pi \tau n^2} e^{2 n \ii
z} \ ,
\label{theta3}
\end{equation}
and satisfies the Jacobi imaginary transformation,
\begin{equation}
\theta_3(z|\tau)=(-\ii \tau)^{-1/2}e^{\ii \tau'2
z^2/\pi} \, \theta_3(z\tau'|\tau') \ ,
\end{equation}
with $\tau'=-1/\tau$. Under the change of variables, $z=\pi r$ and
$\tau=\ii \pi/s$, the $\theta$-function becomes a Gaussian,
which is the relevant function for performing the Ewald sums,
\begin{equation}
\sum_{n=-\infty}^{+\infty}e^{-s(r+n)^2}=(\pi/s)^{1/2}\sum_{n=-\infty}^{+\infty}
e^{-\pi^2 n^2/s}e^{2\pi \ii n r} \ .
\label{1djacobi}
\end{equation}
This expression will be used later, in the derivation of the Ewald sum in
one-dimensional systems.  Equation (\ref{1djacobi}) may be easily generalised to the
2D geometry,
\begin{equation}
\sum_{\bs{n}}e^{-s|\bs{r}+\bs{n}|^2} = (\pi/s)\sum_{\bs{n}}
e^{-\pi^2 n^2/s}e^{2\pi \ii \bs{n}\bs{r}} \ .
\label{2djacobi}
\end{equation}
Comparing this result for 2D with its 1D (\ref{1djacobi}) and
3D(\ref{3djacobi})  counterparts  one finds that the
dimensionality $D$ affects only the constant multiplier as $(\pi/s)^{D/2}$.

\subsection{Derivation}
The analytical derivation of the Ewald sum in 2D proceeds similarly to
the one already presented for 3D. Equations from  (\ref{ep1}) to (\ref{psifinsplit})
are also valid here because their derivation is done without explicit
reference to the dimensionality of the problem. In particular,
the integral $\psi_{\rm inf}(\bs{r},s)$ converges absolutely and to the same value
\begin{equation}
\psi_{{\rm inf}}(\bs{r},s) \lims \frac{1}{\Gamma (\frac{k}{2})}\sum_{\bs{n}}
\frac{\Gamma (\frac{k}{2},\alpha ^2|\bs{r}+\bs{n}|^2 )}{|\bs{r}+\bs{n}|^k}
\ .
\label{i01infpolytropic2d}
\end{equation}
We make the same decomposition of the integral $\psi_{{\rm
fin}}(\bs{r},s)$ as in 3D,
\begin{equation}
\psi_{{\rm fin}}(\bs{r},s) = \psi_{{\rm fin}}^{\bs{n}\neq \bs{0}}(\bs{r},s)
+\psi_{{\rm fin}}^{\bs{n}=\bs{0}}(\bs{r},s) \ ,
\end{equation}
with
\begin{eqnarray}
 \psi_{{\rm fin}}^{\bs{n}\neq \bs{0}}(\bs{r},s)  & = &\frac{\pi}{\Gamma(\frac{k}{2})}
\sum_{\bs{n}\neq \bs{0}} \int_0^{\alpha^2}\frac{t^{\frac{k}{2}-1}}{(t+s)}
\exp\left[\frac{-\pi^2n^2}{t+s}+2\pi \ii
\bs{n}\bs{r}\right]\di t \label{psifinnneq02d}\\
\psi_{{\rm fin}}^{\bs{n}=\bs{0}}(\bs{r},s)& = &\frac{\pi}{\Gamma(\frac{k}{2})}
\int_0^{\alpha^2}\frac{t^{\frac{k}{2}-1}}{(t+s)}
\di t \ ,
\label{i11tildawodiv22d}
\end{eqnarray}
where the two-dimensional variant of the Jacobi
transformation~(\ref{2djacobi}) is used. The
difference between the pair of equations
(\ref{psifinnneq02d}, \ref{i11tildawodiv22d}) and their
three-dimensional
analogues~(\ref{psifinnneq0}, \ref{i11tildawodiv2}) relies in a
substitution of the 3D factor $(\pi/(t+s))^{3/2}$  by  the 2D one $\pi/(t+s)$.

First, we consider the term $\psi_{{\rm fin}}^{\bs{n}\neq \bs{0}}(\bs{r},s)$.
Following the same analysis as for its 3D counterpart, it can be
shown that this parametric integral also converges absolutely. It yields
\begin{eqnarray}
\psi_{{\rm fin}}^{\bs{n}\neq \bs{0}}(\bs{r},s)=&&\frac{\pi}{\Gamma(\frac{k}{2})}
\sum_{\bs{n}\neq \bs{0}}e^{2\pi \ii \bs{n}\bs{r}} \int_{0}^{\alpha^2}t^{\frac{k}{2}-2}
\exp\left[-\frac{\pi^2n^2}{t}\right] \di t \nonumber \\
=&&\sum_{\bs{n}\neq \bs{0}}\frac{\pi\cos(2\pi \bs{n}\bs{r})}{\Gamma(\frac{k}{2})}
\alpha^{k-2} E_{\frac{k}{2}}\left(\frac{\pi ^2n^2}{\alpha ^2}\right) \ .
\label{i01nnot0fin2d}
\end{eqnarray}

The modification of the integral $\psi_{{\rm fin}}^{\bs{n}=\bs{0}}$ is
less straightforward, since it requires specific integrations and
expansions in the series for small $s$. Namely, we have to evaluate the
integral
\begin{equation}
\psi_{{\rm fin}}^{\bs{n}=\bs{0}} = \frac{\pi}{\Gamma(\frac{k}{2})} \int_{s/(\alpha^2+s)}^{1}
\frac{(1-v)^{\frac{k}{2}-1}}{v^{\frac{k}{2}}}s^{k/2-1}\di v
\label{i11n0withexp2D}
\end{equation}
which is the 2D equivalent of Eq.~(\ref{i11n0withexp}).

In the following, we consider separately the cases of long-range potential ($1\leq k<2$),
marginal interaction ($k=2$) and short-range potential ($k>2$).
\begin{enumerate}
\item
$1\leq k<2$. As in 3D, the integral can be found analytically via the
incomplete beta- and incomplete gamma-function with known series expansions
for small $s$. Omitting these unnecessary intermediate expressions, we
give the final expansion for $\psi_{{\rm fin}}^{\bs{n}=\bs{0}}$,
\begin{equation}
\psi_{{\rm fin}}^{\bs{n}=\bs{0}} =s^{\frac{k-2}{2}} \frac{\pi^2}
{\sin\left(\frac{k\pi}{2}\right) \Gamma\left(\frac{k}{2}\right)}+\frac{2\pi \alpha ^{k-2}}
{(k-2) \Gamma\left[\frac{k}{2}\right]}+ \mathcal{O}(s^{k/2})  \ .
\label{i11k122D}
\end{equation}
The first term of the expansion,
\begin{equation}
S(s)=s^{\frac{k-2}{2}}\frac{\pi^2}{\sin\left(\frac{k\pi}{2}\right)
\Gamma\left(\frac{k}{2}\right)} \ ,
\end{equation}
clearly diverges when $s\rightarrow 0$. Similarly to the 3D case, this term is cancelled
in a charge-neutral cell and  hence,
\begin{equation}
\psi_{{\rm fin}}^{\bs{n}=\bs{0}} =\frac{2\pi \alpha ^{k-2}}{(k-2)
\Gamma\left[\frac{k}{2}\right]}
\label{i11k122Dfin}
\end{equation}

\item
$k=2$. The integration of Eq.~(\ref{i11n0withexp2D}) is performed to yield in the
limit $s\rightarrow 0$ a marginal logarithmic dependence,
\begin{equation}
\psi_{{\rm fin}}^{\bs{n}=\bs{0}}=-\pi \ln s + 2\pi \ln \alpha +
\mathcal{O}(s \ln s) \ .
\label{i11k22D}
\end{equation}
As for the 3D geometry, the jellium model is inapplicable in this
particular case since the energy of the continuous background diverges.
Nonetheless the diverging component
\begin{equation}
S(s)=-\pi \ln s
\end{equation}
can be removed if we consider a charge-neutral system with a finite number
of charges. In this case,
\begin{equation}
\psi_{{\rm fin}}^{\bs{n}=\bs{0}}=2\pi \ln \alpha   \ .
\label{i11k22Dfin}
\end{equation}

\item
$k>2$. The integral (\ref{i11n0withexp2D}) can be evaluated by taking $s=0$, since
its convergence is absolute,
\begin{equation}
\psi_{{\rm fin}}^{\bs{n}=\bs{0}} \lims \frac{2 \pi \alpha ^{k-2}}{(k-2) \Gamma\left[\frac{k}{2}\right]}\label{i11km22Dfin}
\end{equation}
\end{enumerate}

The second potential energy component, $I_{00}$ (\ref{ep1}), is calculated
as in the 3D case. The result for 2D is
\begin{eqnarray}
I_{00}(s) & = & \sum_{i=1}^N q_i^2 (\psi_{{\rm fin}}^{\bs{n}\neq\bs{0}}(\bs{0},s)+
\psi_{\rm inf}(\bs{0},s)-\psi_{{\rm inf}}^{\bs{n}=\bs{0}}(\bs{0},s))
\\
& = & \sum_{i=1}^N q_i^2\left[\sum_{\bs{n}}\frac{\Gamma (\frac{k}{2},\alpha ^2n^2 )}
{\Gamma (\frac{k}{2})n^k}+\sum_{\bs{n}\neq \bs{0}}\frac{\pi \alpha^{k-2}}{\Gamma(\frac{k}{2})}
E_{\frac{k}{2}}\left(\frac{\pi ^2n^2}{\alpha ^2}\right)-
\frac{\alpha^k}{\Gamma(\frac{k}{2}+1)} +\psi_{{\rm fin}}^{\bs{n}=\bs{0}} \right] \ . \nonumber
\label{xi2dfin0}
\end{eqnarray}

\subsection{Final expressions}

With respect to the 3D case, the changes in the 2D Ewald sum
appear in those terms in which the Jacobi transformation is used, that is
in $\kappa(\bs{n})$ and $C_1$,
\begin{eqnarray}
\kappa(\bs{n})& = &\frac{\pi \alpha^{k-2}}{\Gamma(\frac{k}{2})} E_{\frac{k}{2}}
\left(\frac{\pi ^2n^2}{\alpha ^2}\right)  \label{defkappa2d} \\
C_1 & = & \psi_{{\rm fin}}^{\bs{n}=\bs{0}}
\end{eqnarray}
The other terms, namely
$R(\bs{r},\bs{n})$, $\rho(\bs{n})$ and $C_2$, are not affected by
dimensionality and may be taken directly from the previous section.

Within the jellium model for a long-range potential ($k<2$), the Ewald sum
is given by
\begin{equation}
\Pi^{{\rm jel}} =
\frac{q_{-}^2}{L^k}\sum_{i<j}\psi(r_{ij}/L)+\frac{Nq_{-}^2}{2L^k}\xi \ .
\label{epgen2dfin}
\end{equation}
A more general form, applicable to any system with a short-range potential
($k>2$), a charge-neutral system with long-range interaction ($k<2$), or a marginal
($k=2$) potential is expressed as
\begin{equation}
\Pi^{{\rm gen}} = \frac{1}{L^k}\sum_{i<j}q_i q_j\psi(r_{ij}/L)+\frac{\xi}{2L^k}
\sum_{i=1}^N q_i^2 \ .
\label{epgen2dfin2}
\end{equation}

In the same way as for the 3D systems we can modify the sum in the
reciprocal space, and with the same notations
(\ref{tildepsi})~--~(\ref{se}) ($\rho$, $R$ and the constants $C_1$, $C_2$
are the new ones, corresponding to 2D case) the potential energy may be
given by
\begin{eqnarray}
\Pi^{{\rm jel}} &&= \frac{q_{-}^2}{L^k}\sum_{i<j}\tilde{\psi}(r_{ij}/L)+\frac{1}{2L^k}
\sum_{\bs{n}\neq 0}\kappa(\bs{n})|\tilde{S}_{\rm equal}|^2+\frac{Nq_{-}^2}{2L^k}\tilde{\xi}\label{epgen2dfineff}\\
\Pi^{{\rm gen}} &&= \frac{1}{L^k}\sum_{i<j}q_i q_j\tilde{\psi}(r_{ij}/L)+\frac{1}{2L^k}
\sum_{\bs{n}\neq 0}\kappa(\bs{n})|\tilde{S_q}|^2+\frac{\sum q_i^2}{2L^k}\tilde{\xi} \
.
\label{epgen2dfin2eff}
\end{eqnarray}
%
%
%
%
%
%
\section{Ewald method for one-dimensional systems}
\label{sec:ewald1d}

As it has been commented before for the 2D case, the differences due to
dimensionality are caused by the form of the Jacobi imaginary
transformation. In the derivation for 1D, one needs the following ones
\begin{eqnarray}
x^{-2s} &=&\frac{1}{\Gamma(s)}\int_0^{\infty}t^{s-1}e^{-tx^2}\di t \label{repx1_1d}\\
\sum_{n=-\infty}^{+\infty}e^{-sn^2}&=&(\pi/s)^{1/2}\sum_{n=-\infty}^{+\infty}e^{-\pi^2 n^2/s}
\label{1djacobi2_n}\\
\sum_{n=-\infty}^{+\infty}e^{-s(r+n)^2}&=&(\pi/s)^{1/2}\sum_{n=-\infty}^{+\infty}
e^{-\pi^2 n^2/s}e^{2\pi \ii n r}  \ .
\label{1djacobi2_rn}
\end{eqnarray}
Similarly to what discussed in the previous section, the only
terms to be changed are those where the Jacobi transformation is used, namely
$\psi_{\rm fin}^{n\neq 0}$ (in $I_{01}$ in a radial-dependent form,
in $I_{00}$ for $r=0$). The difference arises from a different power
exponent ($1/2$) in (\ref{1djacobi2_n}) and (\ref{1djacobi2_rn}), that is in
(\ref{i01nnot0fin}) $k$ has to be substituted by $k+2$ (and $\pi^{3/2}$ --
by $\pi^{1/2}$, respectively), yielding
\begin{equation}
\psi_{\rm fin}^{n\neq 0}=\sum_{n\neq 0}\frac{\pi^{1/2} e^{2\pi \ii nr}}{\Gamma(\frac{k}{2})}
\alpha^{k-1} E_{\frac{k+1}{2}}\left(\frac{\pi ^2 n^2}{\alpha ^2}\right) \ .
\label{i11not01D}
\end{equation}

As far as the term $\psi_{\rm fin}^{n=0}$ is concerned, we should perform a
simple integration and do a series expansion for small $s$,
\begin{equation}
\psi_{\rm fin}^{n=0} = \frac{\pi^{1/2}}{\Gamma(\frac{k}{2})}
\int_{s/(\alpha^2+s)}^{1}\frac{(1-v)^{\frac{k-1}{2}}}{v^{\frac{k+1}{2}}}\di v
\label{i11n0withexp1D}
\end{equation}
The estimation of this integral depends on the $k$ value. In the following,
we detail this analysis.

\begin{enumerate}
\item
$k = 1$, the marginal case,
\begin{equation}
\psi_{\rm fin}^{n=0} =  \frac{\pi^{1/2}}{\Gamma(\frac{k}{2})} (-\ln s -2+2 \ln (2\alpha))+
\mathcal{O}(s) \ .
\label{i11n0k11D}
\end{equation}
As before, we keep only the constant term, considering the diverging term
absent due to the charge neutrality condition. Therefore, with
$\Gamma(1/2)=\sqrt{\pi}$ one has
\begin{equation}
\psi_{\rm fin}^{n=0} =  -2+2 \ln (2\alpha)
\label{i11n0k11Dfin}
\end{equation}
\item
$k>1$, the short-range potential,
\begin{equation}
\psi_{\rm fin}^{n=0} =  \frac{\pi^{1/2}}{\Gamma(\frac{k}{2})} \cdot \frac{2\alpha^{k-1}}{k-1}+
\mathcal{O}(s)+\mathcal{O}(s^{(k-1)/2} \ln s) \ .
\label{i11n0km11D}
\end{equation}
In the limit $s\rightarrow 0$, it yields
\begin{equation}
\psi_{\rm fin}^{n=0} =\frac{2\pi^{1/2}\alpha^{k-1}}{(k-1)\Gamma(\frac{k}{2})}
\label{i11n0km11Dfin}
\end{equation}
resembling the 3D result (\ref{i01kgt3}), with the change $k\rightarrow
k+2$ (except in the $\Gamma$ term) and $\pi^{3/2}\rightarrow \pi^{1/2}$.
\end{enumerate}

The final result for the one-dimensional Ewald summation reads
\begin{eqnarray}
\psi(\bs{r}) & =& \sum_{n}R(n,r)+\sum_{n\neq 0}K(n,r))+C_1\\
\xi & = & \sum_{n\neq 0}(\rho(n)+\kappa(n))+C_1+C_2 \ ,
\end{eqnarray}
where $C_1=\psi_{fin}^{n=0}$ is taken from the
expressions~(\ref{i11n0k11Dfin}) (if $k=1$) or (\ref{i11n0km11Dfin}) (if
$k>1$).

For $k=1$, the only consistent system is the charge-neutral one with a finite number
of particles. In this case and for a short-range potential ($k>1$) one the
potential energy is given by
\begin{equation}
\Pi^{{\rm gen}} = \frac{1}{L}\sum_{i<j}q_i
q_j\psi(r_{ij}/L)+\frac{\sum_{i=1}^{N} q_i^2}{2L^k}\xi \ .
\label{epgen1dfin}
\end{equation}

Although the Ewald method is applicable to one-dimensional
problems, there is a direct way to calculate the sums for polytropic
potentials
\begin{equation}
\Pi = \frac{1}{L^k}\sum_{n=-\infty}^{n=+\infty}\frac{1}{|r+n|^k} \ .
\label{ep1ddirect}
\end{equation}
For $k>1$, this sum can be represented as
a linear combination of the Hurwitz zeta functions,
\begin{equation}
\frac{1}{L^k}\sum_{n=-\infty}^{+\infty}\frac{1}{|r+n|^k}=\frac{1}{L^k}(H_k(r)+H_k(1-r))
\ .
\label{ep1ddirect2}
\end{equation}
In particular, for $k=2$ the sum converts into a familiar expression used
in the Calogero-Sutherland model~\cite{Sutherland1971,Astra2006},
\begin{equation}
\frac{1}{L^2}\sum_{n=-\infty}^{+\infty}\frac{1}{|r+n|^2}=\frac{\pi^2}{L^2 \sin^2(\pi r) } \ .
\label{ep1ddirect3}
\end{equation}
Notice that the sum (\ref{ep1ddirect2}) may be expressed in terms of
trigonometric functions only for even values of $k$ via $(k-2)$ times
differentiation of Eq. (\ref{ep1ddirect3}). Anyway, the possibility to find
exact expressions for infinite sums in 1D suggests that the use of the
Ewald method might not be needed, but we keep it as a possibly useful mathematical
relation and for completeness.
%
%
\section{Ewald method in a rectangular box of arbitrary side lengths}
\label{sec:ewaldbox}

\subsection{3D case}

A special and interesting situation arises if we consider a simulation cell in a more
general way, as a rectangular box with different side lengths
($L_x,\:L_y,\:L_z$ in the corresponding dimensions). The need to deal with
a box of unequal size lengths may occur in the simulation of a solid with
a noncubic lattice (the simplest examples include a hexagonal closed packed
crystal in 3D geometry), since the lattice vectors $\bs{n}$ in the sum over
images on (\ref{initialhamiltonian1}) are no longer orthogonal.
Focusing our analysis to a 3D geometry, the potential energy is now given
by
\begin{equation}
\Pi=\frac{1}{2}{\sum_{\bs{n_a} \in \mz^3}} ^{\prime} \left[\sum_{i=1}^{N}
\sum_{j=1}^{N}\phi(\bs{r}_{ij}+L_0\bs{n}_r)\right] \ ,
\label{initialhamiltonian1nbox}
\end{equation}
with $\bs{n}_r=(\bs{n}_xL_x+\bs{n}_yL_y+\bs{n}_zL_z)/L_0$,
$\bs{n}_{x,y,z}$ being integer vectors along the corresponding
axis $x,\:y,\:z$. We have introduced the geometric average
$L_0=(L_xL_yL_z)^{1/3}$ and we will use reduced $L_0$ units for $r_{ij}$, and hence
$r_{ij}$ will be adimensional.   Repeating the standard procedure,
we multiply the potential energy by a Gaussian term
$\exp(-s|\bs{n}_r+\bs{r}|^2)$ and, at the end, we take the limit $s\rightarrow 0$,
separating the converging part, if present. We group separately the
interaction with images of other particles $I_{01}$ and the interaction of
a particle with its own images $I_{00}$,
\begin{equation}
\Pi = \frac{1}{L_0^k}(I_{01} + I_{00}) \ ,
\label{ep1nbox}
\end{equation}
where
\begin{eqnarray}
I_{01} & = & \sum_{\bs{n} \in \mz^3} \left[\sum_{1\leq i<j\leq N}
\frac{q_i q_j e^{-s |\bs{n}_r+\bs{r}_{ij}|^2}}{|\bs{r}_{ij}+\bs{n}_r|^k}\right]
\label{i01initnbox}\\
I_{00} & =& \frac{1}{2}\sum_{\bs{n} \in \mz^3\backslash \bs{0}}
\frac{e^{-s |\bs{n}_r^2|}}{|\bs{n}_r|^k}\sum_{i=1}^{N}q_i^2 \ .
\label{i00initnbox}
\end{eqnarray}
Comparing the relations (\ref{ep1nbox})~--~(\ref{i00initnbox}) to the cubic
case (\ref{ep1})~--~(\ref{i00init}), one notices that these relations
remain  unchanged if $\bs{n}$ is formally substituted by $\bs{n}_r$, and
the constant coefficient $1/L^k$ is replaced by $1/L_0^k$. Therefore, all
the results found without the Jacobi transformation (\ref{3djacobi}) remain the
same with $\bs{n}_r$ instead of $\bs{n}$. In particular,
Eq.~(\ref{i01infpolytropic}) transforms into the following
\begin{equation}
\psi_{{\rm inf}}= \frac{1}{\Gamma (\frac{k}{2})}\sum_{\bs{n}}
\frac{\Gamma (\frac{k}{2},\alpha ^2|\bs{r}+\bs{n}_r|^2
)}{|\bs{r}+\bs{n}_r|^k} \ .
\label{i01infpolytropicnbox}
\end{equation}
The Jacobi transformation (\ref{3djacobi}) in a noncubic box has the following form
\begin{eqnarray}
\sum_{\bs{n}_r} e^{-s|\bs{n}_r+\bs{r}|^2}=&&\prod_{i=x,y,z}\sum_{n_i} e^{-s(n_iL_i/L_0+r_i)^2}
\nonumber\\
=&& \left[ \prod_{i=x,y,z}\left(\frac{\pi}{s(L_i/L_0)^2}\right)^{1/2}
\right] \prod_{i=x,y,z}\sum_{n_i}\exp\left(-\frac{\pi^2n_i^2}{s(L_i/L_0)^2}\right)
\exp(2\pi \ii n_i r_iL_0/L_i )\nonumber\\
=&&(\pi/s)^{3/2}\sum_{\bs{n}_k}\exp(-\pi^2 |\bs{n}_k|^2/s) \exp(2\pi \ii
\bs{n}_k \bs{r} ) \ ,
\label{3djacobinbox}
\end{eqnarray}
with $\bs{n}_k=\bs{n}_x L_0/L_x+\bs{n}_y L_0/L_y+\bs{n}_z L_0/L_z$
the normalised displacement vector in momentum space.  The last equation is
obtained from the original expression (\ref{3djacobi}) by a formal
substitution of the vector $\bs{n}$ by $\bs{n}_k$.

In order to calculate $\psi_{\rm fin}$ we first modify Eq.~(\ref{transf1}),
\begin{equation}
\exp[-s|\bs{n}_r+\bs{r}|^2-t|\bs{n}_r+\bs{r}|^2]=\exp\left[-(s+t)|\bs{n}_r+\bs{r}|^2\right]
\ ,
\label{transf1nbox}
\end{equation}
then insert it into the relation~(\ref{3djacobinbox}), and finally separate
the summand $\bs{n}=\bs{0}$,
\begin{eqnarray}
\psi_{\rm fin} &= & \frac{\pi^{\frac{3}{2}}}{\Gamma(\frac{k}{2})}\sum_{\bs{n}_k\neq \bs{0}}
\int_0^{\alpha^2}\frac{t^{\frac{k}{2}-1}}{(t+s)^{\frac{3}{2}}}
\exp\left[\frac{-\pi^2\bs{n}_k^2}{t+s}+2\pi \ii \bs{n}_k\bs{r}\right]\di t+
\frac{\pi^{\frac{3}{2}}}{\Gamma(\frac{k}{2})}\int_0^{\alpha^2}\frac{t^{\frac{k}{2}-1}}
{(t+s)^{\frac{3}{2}}} \di t
 \nonumber\\
&= & \psi_{\rm fin}^{\bs{n}\neq \bs{0}}+\psi_{\rm fin}^{\bs{n}=\bs{0}} \ .
\label{i11tildawodiv2nbox}
\end{eqnarray}
The subsequent derivation follows exactly the
derivation for a cubic box, with the change of $\bs{n}$ by
$\bs{n}_r$  and  $\bs{n}_k$  for  sums
in the real and momentum spaces, respectively. The final
result for a 3D system in a noncubic box can be summarised as follows
\begin{eqnarray}
\psi(\bs{r})=&&\sum_{\bs{n}_r}\frac{\Gamma(k/2,\alpha^2|\bs{n}_r+\bs{r}|^2)}
{\Gamma(k/2)|\bs{n}_r+\bs{r}|^k}
+\sum_{\bs{n}_k\neq \bs{0}}\frac{\pi^{\frac{3}{2}}\alpha^{k-3}\cos(2\pi
\bs{n}_k \bs{r})}{\Gamma(k/2)}E_{\frac{k-1}{2}}\left(\frac{\pi^2|\bs{n}_k|^2}{\alpha^2}\right)
+C_1\label{psinbox}\\
\xi=&&\sum_{\bs{n}_r\neq \bs{0}}\frac{\Gamma(k/2,\alpha^2|\bs{n}_r|^2)}
{\Gamma(k/2)|\bs{n}_r|^k}
+\sum_{\bs{n}_k\neq \bs{0}}\frac{\pi^{\frac{3}{2}}\alpha^{k-3}}{\Gamma(k/2)}E_{\frac{k-1}{2}}\left(\frac{\pi^2|\bs{n}_k|^2}
{\alpha^2}\right)
+C_1+C_2\label{xinbox}\\
\Pi = &&\frac{q_{-}^2}{L_0^k}\sum_{i<j}\psi(r_{ij}/L_0)+\frac{Nq_{-}^2}{2L_0^k}\xi\label{epgenfinnbox}
\end{eqnarray}
with the constants $C_1$ and $C_2$ defined in (\ref{c1}) and (\ref{c2}). As
it was done in the cubic box, the potential energy may also be given with the momentum
space sum (linear in $N$). Applying the definitions, similar to
Eqs (\ref{tildepsi})~--~(\ref{se}),
\begin{eqnarray}
\tilde{\psi}(\bs{r})&&=\sum_{\bs{n}_r}R(\bs{n}_r,\bs{r})+C_1\label{tildepsinoncubic}\\
\tilde{\xi}&&=\sum_{\bs{n}_r\neq \bs{0}}\rho(\bs{n}_r)+C_1+C_2\label{tildexinoncubic}\\
\tilde{S}_{\rm equal}(\bs{n}_k)&&=q_{-}\sum_{j}\exp(2\pi \ii \bs{n}_k\bs{r}_j/L)\label{sqnoncubic}\\
\tilde{S_q}(\bs{n}_k)&&=\sum_{j}q_j\exp(2\pi \ii
\bs{n}_k\bs{r}_j/L) \ ,
\label{senoncubic}
\end{eqnarray}
the potential energy for a one-component jellium model converts into
\begin{equation}
\Pi^{\rm jel} = \frac{q_{-}^2}{L_0^k}\sum_{i<j}\tilde{\psi}(r_{ij}/L_0)+\frac{1}{2L_0^k}
\sum_{\bs{n}_k\neq 0}\kappa(\bs{n}_k)|\tilde{S}_{\rm equal}(\bs{n}_k)|^2+\frac{Nq_{-}^2}{2L_0^k}\tilde{\xi} \
,
\label{epjelfinnboxeff}
\end{equation}
with a natural extension to the general case
\begin{equation}
\Pi^{\rm gen} = \frac{1}{L_0^k}\sum_{i<j}q_i q_j\tilde{\psi}(r_{ij}/L_0)+
\frac{1}{2L_0^k}\sum_{\bs{n}_k\neq 0}\kappa(\bs{n}_k)|\tilde{S_q}(\bs{n}_k)|^2+\frac{\sum
q_i^2}{2L_0^k}\tilde{\xi} \ .
\label{epgenfinnboxeff}
\end{equation}

\subsection{2D case}

The generalization of the formulae found in a square 2D geometry to a rectangular
simulation box comes in a similar manner. It is sufficient to take the
resulting expressions for the two-dimensional problem~(\ref{i01infpolytropic2d})
and~(\ref{i01nnot0fin2d}), and to perform the necessary substitutions
$\bs{n}\rightarrow\bs{n}_r$ and $\bs{n}\rightarrow\bs{n}_k$,
\begin{eqnarray}
\psi(\bs{r})=&&\sum_{\bs{n}_r}\frac{\Gamma(k/2,\alpha^2|\bs{n}_r+\bs{r}|^2)}{\Gamma(k/2)
|\bs{n}_r+\bs{r}|^k}
+\sum_{\bs{n}_k\neq \bs{0}}\frac{\pi\alpha^{k-2}\cos(2\pi  \bs{n}_k \bs{r})}
{\Gamma(k/2)}E_{\frac{k}{2}}\left(\frac{\pi^2|\bs{n}_k|^2}{\alpha^2}\right)
+\psi_{\rm fin}^{\bs{n}=\bs{0}} \\
\xi=&&\sum_{\bs{n}_r\neq \bs{0}}\frac{\Gamma(k/2,\alpha^2|\bs{n}_r|^2)}
{\Gamma(k/2)|\bs{n}_r|^k}
+\sum_{\bs{n}_k\neq \bs{0}}\frac{\pi\alpha^{k-2}}{\Gamma(k/2)}E_{\frac{k}{2}}\left(\frac{\pi^2|\bs{n}_k|^2}
{\alpha^2}\right)
+ \psi_{\rm fin}^{\bs{n}=\bs{0}} - \frac{\alpha^k}{\Gamma(\frac{k}{2}+1)} \
,
\end{eqnarray}
where $\psi_{\rm fin}^{\bs{n}=\bs{0}}$ is given by the
expressions~(\ref{i11k122Dfin}), (\ref{i11k22Dfin}) or (\ref{i11km22Dfin}).

For a long-range interaction within the jellium model, the potential energy
becomes
\begin{equation}
\Pi^{\rm jel} = \frac{q_{-}^2}{L_0^k}\sum_{i<j}\psi(r_{ij}/L_0)+\frac{Nq_{-}^2}{2L_0^k}\xi
\label{epgen2dfinnbox1}
\end{equation}
with the notation
\begin{eqnarray}
L_0 & =& (L_xL_y)^{1/2}\label{l02d}\\
\bs{n}_r& =& \bs{n}_xL_x/L_0+\bs{n}_yL_y/L_0\label{nl2d}\\
\bs{n}_k &= & \bs{n}_xL_0/L_x+\bs{n}_yL_0/L_y \ .
\label{nr2d}
\end{eqnarray}
For a multicomponent gas (quasi-neutral in case of a long-range potential),
the potential energy is
\begin{equation}
\Pi^{\rm gen} = \frac{1}{L_0^k}\sum_{i<j}q_i q_j\psi(r_{ij}/L_0)+\frac{\sum
q_i^2}{2L_0^k}\xi \ .
\label{epgen2dfinnbox2}
\end{equation}

Finally, the usual modification to calculate the momentum space sum linearly in $N$
is given by
\begin{eqnarray}
\Pi^{\rm jel}&& = \frac{q_{-}^2}{L_0^k}\sum_{i<j}\tilde{\psi}(r_{ij}/L_0)+\frac{1}{2L_0^k}
\sum_{\bs{n}_k\neq 0}\kappa(\bs{n}_k)|\tilde{S}_{\rm equal}(\bs{n}_k)|^2+\frac{Nq_{-}^2}{2L_0^k}\tilde{\xi}\\
\Pi^{\rm gen}&& = \frac{1}{L_0^k}\sum_{i<j}q_i q_j\tilde{\psi}(r_{ij}/L_0)+\frac{1}{2L_0^k}
\sum_{\bs{n}_k\neq 0}\kappa(\bs{n}_k)|\tilde{S_q}(\bs{n}_k)|^2+\frac{\sum
q_i^2}{2L_0^k}\tilde{\xi} \ ,
\end{eqnarray}
with $\tilde{\psi},\:\tilde{\xi},\:\tilde{S}_{\rm equal},\:\tilde{S_q}$ defined by
(\ref{tildepsinoncubic})~--~(\ref{senoncubic}) in their corresponding
two-dimensional variants.
%
%
%
%
%
\section{Equation summary}
\label{sec:results}
In the previous sections, we have derived general expressions of the Ewald
sums for polytropic $1/|\bs{r}|^k$ potentials in three- two- and
one-dimensional systems. For
integer values of $k$, the polytropic potential reduces to a power-law
interaction, which comprises realizations of high physical relevance.
Integer power-law potentials include
\begin{itemize}
	\item $k=1$ -- Coulomb $1/|\bs{r}|$ interaction;
	\item $k=2$ -- Calogero-Sutherland $1/|\bs{r}|^2$ interaction;
	\item $k=3$ -- dipole-dipole $1/|\bs{r}|^3$ interaction;
	\item $k=4,\:5,\:6$ -- interaction between different Rydberg atoms;
	\item $k=6,\:12$ -- Van der Waals interaction.
\end{itemize}

The expressions for the potential energy for both the jellium model and
the general case of a charge-neutral simulation cell are the following
\begin{eqnarray}
\Pi^{{\rm gen}} =&& \frac{1}{L_0^k}\sum_{i<j}q_i q_j\psi(r_{ij}/L_0)+\frac{\sum q_i^2}{2L_0^k}\xi\\
\Pi^{{\rm jel}} =&& \frac{q_{-}^2}{L_0^k}\sum_{i<j}\psi(r_{ij}/L_0)+\frac{Nq_{-}^2}{2L_0^k}\xi\\
\psi(\bs{r})=&& \sum_{\bs{n}}R(\bs{n}_r,\bs{r})+\sum_{\bs{n}\neq 0}K(\bs{n}_k,\bs{r})+C_1\\
\xi =&& \sum_{\bs{n}\neq 0}(\rho(\bs{n}_r)+\kappa(\bs{n}_k))+C_1+C_2\\
R(\bs{n},\bs{r})=&&\rho(\bs{n}+\bs{r})\\
K(\bs{n},\bs{r})=&&\kappa(\bs{n})\cos(2\pi\bs{n}\bs{r})\\
C_{1}^{{\rm 3D}}=&&\left\{
\begin{array}{ll}
\frac{2 \pi ^{\frac{3}{2}} \alpha ^{k-3}}{(k-3) \Gamma\left[\frac{k}{2}\right]} & \textrm{if $k \neq 3$}\\
-4\pi+4\pi\ln(2\alpha) & \textrm{if $k=3$}\\
\end{array}\right.\\
C_{1}^{{\rm 2D}}=&&\left\{
\begin{array}{ll}
\frac{2\pi \alpha ^{k-2}}{(k-2) \Gamma\left[\frac{k}{2}\right]} & \textrm{if $k \neq 2$}\\
2\pi\ln(\alpha) & \textrm{if $k=2$}\\
\end{array}\right.\\
C_{2}=&&-\frac{\alpha^k}{\Gamma(\frac{k}{2}+1)}\\
L_0=&&\left\{
\begin{array}{ll}
(L_xL_yL_z)^{1/3} & \textrm{in 3D}\\
(L_xL_y)^{1/2} & \textrm{in 2D}\\
\end{array}\right.\\
\bs{n}_r=&&(\bs{n}\cdot\bs{L})/L_0,\textrm{ with }\bs{L}=(L_x,L_y,L_z)\\
\bs{n}_k=&&(\bs{n}\cdot\bs{L}')L_0,\textrm{ with
}\bs{L}'=(1/L_x,1/L_y,1/L_z) \ .
\end{eqnarray}
Alternatively, by performing a momentum space sum the above set of
equations become
\begin{eqnarray}
\Pi^{\rm gen} =&& \frac{1}{L_0^k}\sum_{i<j}q_i q_j\tilde{\psi}(r_{ij}/L_0)+\frac{1}{2L_0^k}\sum_{\bs{n}_k\neq 0}\kappa(\bs{n}_k)|\tilde{S_q}(\bs{n}_k)|^2+\frac{\sum q_i^2}{2L_0^k}\tilde{\xi}\label{pigeneff} \\
\Pi^{\rm jel} =&& \frac{q_{-}^2}{L_0^k}\sum_{i<j}\tilde{\psi}(r_{ij}/L_0)+\frac{1}{2L_0^k}\sum_{\bs{n}_k\neq 0}\kappa(\bs{n}_k)|\tilde{S}_{\rm equal}(\bs{n}_k)|^2+\frac{Nq_{-}^2}{2L_0^k}\tilde{\xi}\label{pijeleff}\\
\tilde{\psi}(\bs{r})&&=\sum_{\bs{n}}R(\bs{n}_r,\bs{r})+C_1\\
\tilde{\xi}&&=\sum_{\bs{n}\neq \bs{0}}\rho(\bs{n}_r)+C_1+C_2\\
\tilde{S}_{\rm equal}&&=q_{-}\sum_{j}\exp(2\pi \ii \bs{n}_k\bs{r}_j/L)\\
\tilde{S_q}&&=\sum_{j}q_j\exp(2\pi \ii \bs{n}_k\bs{r}_j/L) \ .
\end{eqnarray}

In accordance with considerations discussed in preceding sections, the simulation cell
has to fulfill the charge neutrality condition ($\sum_{i=1}^N q_i=0$) for
long-range potentials. Also, notice that in the particular case of a cubic
simulation cell, $\bs{n}_r=\bs{n}_k=\bs{n}$.

Explicit expressions of the coefficients $\rho(\bs{n})$ and
$\kappa(\bs{n})$ for the most relevant interactions are summarised
for 3D and 2D systems in 
Table~1 and Table~2,
respectively.

\begin{table}[h!b!p!]
\tabnote{Table 1. Coefficients $\rho(\bs{n})$ and $\kappa(\bs{n})$ taken from Eqs~(\ref{defrho}) and (\ref{i01nnot0fin}) for 3D geometry. LR and SR stand for long range and short range, respectively.}
\begin{tabular*}{0.95\textwidth}{@{\extracolsep{\fill}} | l | l | l |}
\hline
          & $\rho(\bs{n})$ & $\kappa(\bs{n})$  \\ \hline
LR $\frac{1}{|r|}$&$\frac{\erfc (\alpha |\bs{n}|)}{|\bs{n}|}$ &
$\frac{1}{\pi n^2}  e^{-\frac{\pi^2 n^2}{\alpha^2}}   $ \\ \hline
LR $\frac{1}{|r|^2}$&$\frac{e^{-\alpha^2 n^2}}{n^2}$ &
$\frac{\pi}{|\bs{n}|}  \erfc \frac{\pi
|\bs{n}|}{\alpha}  $ \\ \hline
SR $\frac{1}{|r|^4}$&$\frac{\alpha^2 n^2+1}{n^4} e^{-\alpha^2 n^2}$ &
$2\pi\left(\sqrt{\pi}\alpha
e^{-\frac{\pi^2 n^2}{\alpha^2}}-\pi^2|\bs{n}|\erfc \frac{\pi
|\bs{n}|}{\bs{\alpha}}\right)$ \\ \hline
SR $\frac{1}{|r|^5}$&$\frac{\erfc(\alpha
|\bs{n}|)}{|\bs{n}|^5}+\frac{4e^{-\alpha^2 n^2}}
{3\sqrt{\pi} |\bs{n}|^5 }(\frac{3\alpha |\bs{n}|}{2}+(\alpha |\bs{n}|)^{3})$ &
$\frac{4\pi \alpha^2}{3}\left(e^{-\frac{\pi^2
n^2}{\alpha^2}}-\frac{\pi^2 n^2}{\alpha^2}E_1(\frac{\pi^2
n^2}{\alpha^2})\right)$ \\ \hline
SR
$\frac{1}{|r|^6}$&$(\frac{\alpha^4}{2n^2}+\frac{\alpha^2}{n^4}+\frac{1}
{n^6})e^{-\alpha^2n^2}$ &
$\frac{\pi^{3/2}\alpha^3}{3}\left( e^{-\frac{\pi^2 n^2}
{\alpha^2}}(1-\frac{2\pi^2 n^2}{\alpha^2})+\frac{2\pi^{7/2}
|\bs{n}|^3}{\alpha^3}\erfc
\frac{\pi |\bs{n}|}{\alpha}\right) $ \\ \hline
SR $\frac{1}{|r|^{12}}$ & $\displaystyle\sum\limits_{m=0}^5\frac{(\alpha
n)^{2m}}{m!}\frac{e^{-\alpha^2n^2}}{n^{12}}$ &
$-\displaystyle\sum\limits_{m=0}^4(-2)^{m+1} (7-2m)!!\left(\frac{\pi
n}{\alpha}\right)^{2m}\frac{e^{-\frac{\pi^2 n^2}{\alpha^2}}}{945}$ \\
 & & $-\frac{32\sqrt{\pi}\left(\frac{\pi
n}{\alpha}\right)^{9}}{945}\erfc \frac{\pi n}{\alpha} $ \\ \hline
\end{tabular*}
\label{table3d}
\end{table}

\begin{table}[h!b!p!]
\tabnote{Table 2. The coefficients $\rho(\bs{n})$ and $\kappa(\bs{n})$ taken from Eqs~(\ref{defrho}) and (\ref{defkappa2d}) for 2D geometry. LR and SR stand for long range and short range, respectively.}
  \begin{tabular*}{0.95\textwidth}{@{\extracolsep{\fill}} | l | l | l |}
\hline
          & $\rho(\bs{n})$ & $\kappa(\bs{n})$  \\ \hline
LR $\frac{1}{|r|}$&  $\frac{\erfc (\alpha |\bs{n}|) }{|\bs{n}|}$ &
$\frac{1}{|\bs{n}|}  \erfc \frac{\pi |\bs{n}|}{\alpha} $ \\ \hline
SR $\frac{1}{|r|^3}$&  $\frac{2\alpha}{\sqrt{\pi}n^2}e^{-\alpha^2n^2}+
\frac{\erfc (\alpha |\bs{n}|) }{|\bs{n}|^3}$ &  $4\left(\sqrt{\pi}\alpha
e^{-\frac{\pi^2 n^2}{\alpha^2}}-\pi^2|\bs{n}|\erfc \frac{\pi
|\bs{n}|}{\bs{\alpha}}\right)$ \\ \hline
SR $\frac{1}{|r|^4}$&  $\frac{\alpha^2
n^2+1}{n^4} e^{-\alpha^2 n^2}$ &
$\pi\alpha^2\left(e^{-\frac{\pi^2 n^2}{\alpha^2}}- \frac{\pi^2
n^2}{\alpha^2} E_1(\frac{\pi^2n^2}{\alpha^2}) \right)$ \\ \hline
SR $\frac{1}{|r|^5}$&  $\frac{\erfc (\alpha |\bs{n}|) }{|\bs{n}|^5}+
\frac{4e^{-\alpha^2n^2}}{3\sqrt{\pi}|\bs{n}|^5}(\frac{3\alpha|\bs{n}|}{2}+(\alpha|\bs{n}|)^{3})$
&
$\frac{8\sqrt{\pi}\alpha^3}{9}\left( e^{-\frac{\pi^2
n^2}{\alpha^2}}(1-\frac{2\pi^2 n^2}
{\alpha^2})+\frac{2\pi^{7/2} |\bs{n}|^3}{\alpha^3}\erfc \frac{\pi
|\bs{n}|}{\alpha}\right)$ \\ \hline
SR $\frac{1}{|r|^6}$&
$(\frac{\alpha^4}{2n^2}+\frac{\alpha^2}{n^4}+\frac{1}
{n^6})e^{-\alpha^2n^2}$ &
$\frac{\pi\alpha^4}{4}\left(e^{-\frac{\pi^2 n^2}{\alpha^2}} (1 -
\frac{\pi^2 n^2}{\alpha^2}) +
  \frac{\pi^4 n^4}{\alpha^4} E_1(\frac{\pi^2 n^2}{\alpha^2})\right)$ \\
  \hline
SR $\frac{1}{|r|^{12}}$&  $\displaystyle\sum\limits_{m=0}^5\frac{(\alpha
n)^{2m}}{m!}\frac{e^{-\alpha^2n^2}}{n^{12}}$ &
$
\displaystyle\sum\limits_{m=0}^4(-1)^m (4-m)!\left(\frac{\pi
n}{\alpha}\right)^{2m}\frac{e^{-\frac{\pi^2 n^2}{\alpha^2}}}{120}$\\
 &    & $-\frac{\left(\frac{\pi
n}{\alpha}\right)^{10}}{120}E_1\left(\frac{\pi^2
n^2}{\alpha^2}\right)$\\  \hline
\end{tabular*}
\label{table2d}
\end{table}

%
%
%
%
\section{Practical application and optimisations in the Ewald technique}
\label{sec:practical}
\subsection{General notes}

The basic idea of the Ewald method is to calculate slowly decaying sums in
a rapid manner by means of the Fourier transform of the slowly converging
part. Although conceptually it provides an exact result, the number of terms
which has to be summed in order to reach the needed convergence  is {\it a priori}
unknown.  Once we choose the interaction potential, this fixes the exact
form of the sums to calculate, and the practical remaining question is the
proper choice of the free parameter $\alpha$ and the numbers of terms to
be calculated in both sums: $N_r$ and $N_k$ in coordinate and momentum spaces,
respectively. The computer time $T$ is a function of
only $N_r$ and $N_k$, $T=t_r N_r+t_k N_k$, with the constants
$t_r$ and $t_k$ depending on the complexity of the coefficients in the sums.
One can notice that $t_k$ is usually much less then $t_r$, since in the
Jacobi-transformed sum we only calculate cosine functions, which is
generally far less time-consuming than the complicated functions appearing in
$R$. It is clear that the parameter $\alpha$ affects only the resulting error
in the energy. In fact, the value of $\alpha$ being very small or very
large eliminates errors in one of the sums, but amplifies them in the other,
so there is an ``optimal'' point for $\alpha$, yielding a minimum error in
the total energy.

In the following, we discuss a way for error ($\delta E$) minimization assuming the
calculation time $T$ fixed. From our point of view, a useful  approach for
practical implementation is represented by the following scheme
\begin{itemize}
\item
We determine a time law $T=t_r N_r+t_k N_k$ in a preliminary calculation
and fix the values of $t_r$ and $t_k$.
\item
We take a set of configurations, corresponding to the equilibrated
state using an initial Ewald summation. Then, we calculate the
exact energies $E_{\rm ex}$ (as a converged result of the Ewald
summation) and the energies $E(\alpha,N_r,N_k)$ biased by a choice of $N_r$
and $N_k$.  For each pair $(N_r,\,\,N_k)$, we find an optimal value of
$\alpha=\alpha_{\rm opt}(N_r,N_k)$.
\item
We choose the goal accuracy $\delta E_{\rm acc}$ (normally, well below the statistical
error). We plot the error as a function
of the computer time spent and choose the less time consumption  case among
the points that lie
below $\delta E_{\rm acc}$, therefore obtaining all the parameters required:
$\alpha$, $N_r$ and $N_k$. From now on, these parameters are used in actual
simulations.
\end{itemize}

\subsection{Example of optimisation}

Let us illustrate the scheme proposed in the preceding subsection taking as
an example the
problem of two-dimensional zero-temperature Bose gas of particles,
interacting through the  $1/|\bs{r}|^3$ potential. The model corresponds to the
dipole-dipole interaction with all dipole moments aligned perpendicularly
to the plane of motion. To describe  the ground-state properties of the system
we use the variational Monte Carlo (VMC) method and a Jastrow wave function
with a two-body correlation factor which is solution of the
two-body scattering problem~\cite{McMillan1965a}.

\begin{figure}[tbp]
\begin{center}
\includegraphics*[width=0.7\columnwidth]{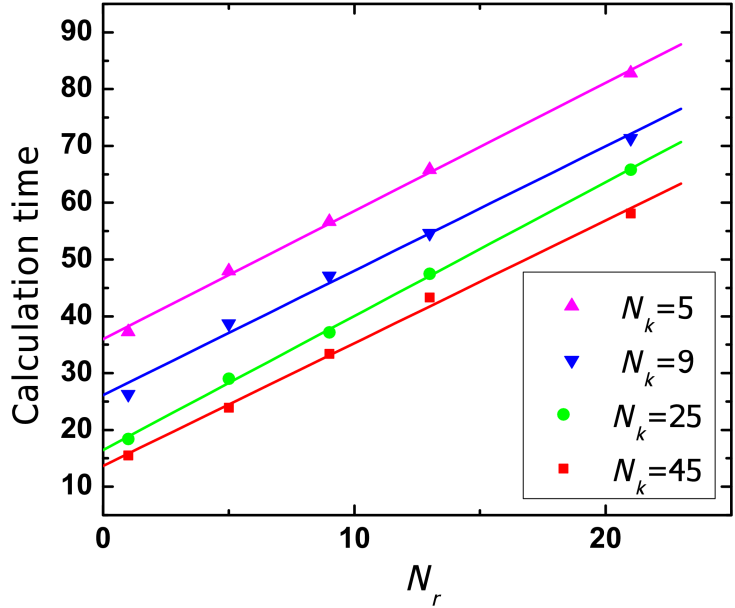}
\end{center}
\caption{Dependence of the calculation time  $T$ on the number of terms $N_r$ in
the coordinate space for fixed numbers of terms in the momentum space $N_k=5,9,25,45$.}
\label{tonnr}
\end{figure}
\begin{figure}[tbp]
\begin{center}
\includegraphics*[width=0.7\columnwidth]{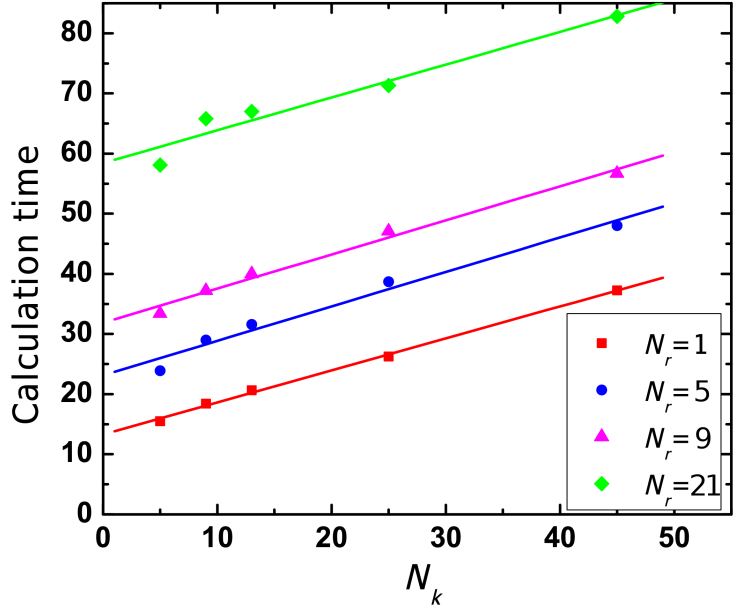}
\end{center}
\caption{Dependence of the calculation time $T$ on the number of terms $N_k$ in
the momentum space for fixed numbers of terms in the coordinate space $N_r=1,5,9,21$.}
\label{tonnk}
\end{figure}

The optimisation is done by averaging over $N_{\rm conf}=50$ uncorrelated VMC
configurations, sampled according to the chosen probability distribution.
We define the error $\delta E(\alpha)$ as a sum over $N_{\rm conf}$
configurations of the difference of the Ewald energy $E(i_{\rm
conf},\alpha,N_r,N_k)$, calculated for a given set of parameters
$(\alpha,N_r,N_k)$ and the converged energy $E_{\rm ex}(i_{\rm conf})=\lim_{N_k
\rightarrow \infty}\lim_{N_r \rightarrow \infty}E(i_{\rm
conf},\alpha,N_r,N_k)$. The dependence of the computer time $T$, needed for the
evaluation of Ewald sums, on the parameter set is shown in Figs~\ref{tonnr} and
\ref{tonnk}. In
Fig.~\ref{tonnr}, we show the dependence of $T$ on the number of terms
$N_r$ in real space for different fixed numbers of terms
$N_k$ in the momentum space. The computation time is proportional to the
number of terms and the resulting dependence is linear in $N_r$. A fixed
number of terms $N_k$ requires a certain amount of calculations which
results in a constant shift. Similarly, keeping $N_r$ fixed and varying
$N_k$ produces a linear dependence in $N_k$ with a constant shift which
depends on $N_r$, as shown in Fig.~\ref{tonnk}.

As one sees in Figs.\ref{tonnr} and \ref{tonnk}, the time dependence is linear both on
$N_k$ and $N_r$,
although the point corresponding to (0,0) in $(N_r,\,N_k)$ does not
necessarily gives $T=0$, since the reported time also contains some
initializing calculations.
The total error in the potential, as it is defined above, is given by
\begin{equation}
\delta E(\alpha)=\sqrt{\sum_{i_{\rm{conf}}=1}^{N_{\rm{conf}}}
\frac{\left(E(\alpha,i_{\rm{conf}})-E_{\rm{ex}}(i_{\rm{conf}})\right)^2}{N_{\rm{conf}}}}
\ .
\label{dalpha1}
\end{equation}
According to our previous considerations, in the
case of very small or very large values of $\alpha$ the error coming from one of
two sums, that is in the real or momentum space, grows and dominates over
the error coming from the other sum; for a certain ``optimal'' range of
$\alpha$ these two errors are of the same order. Notice that for each
particular configuration, and each pair $(N_r,\,N_k)$, it is possible to find
$\alpha_{\rm opt}(i_{conf})$, such that
$E(\alpha_{\rm opt}(i_{conf}),i_{conf})-E_{\rm ex}(i_{conf})=0$. Instead, our task
is to obtain a ``universal'' parameter $\alpha_0$, minimizing the total
error~(\ref{dalpha1}). The mean over the configuration set of the biased
energies $\bar{E}(\alpha_0, i_{\rm{conf}})$ is used as an estimation for
the mean of the exact energies $\bar{E}_{ex}$, introducing an inevitable
systematic error.  As it appears in typical calculations, this error is at
least one order of magnitude smaller than the statistical error~(\ref{dalpha1}) given by
the minimization of $\delta E(\alpha)$.

\begin{figure}[tbp]
\begin{center}
\includegraphics*[width=0.6\columnwidth]{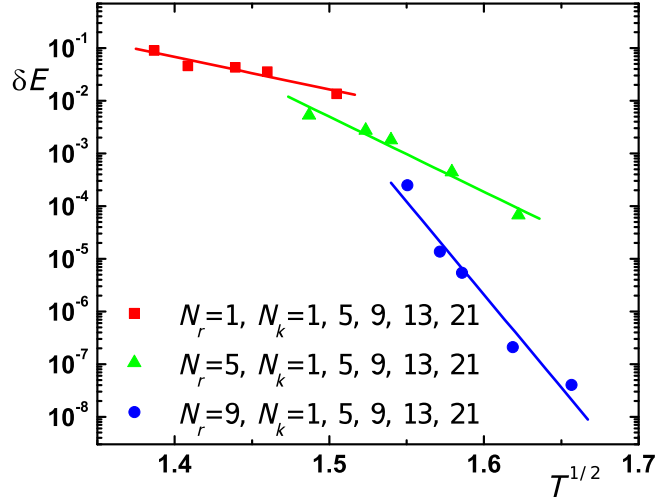}
\end{center}
\caption{Resulting error as a function of the computer time for different parameter sets.}
\label{erront4}
\end{figure}

A second step is the study of the dependence of the error and time on different pairs
$(N_r,\,N_k)$. The calculation time can be split as the sum of times for
summing up in real and momentum spaces,
\begin{equation}
T=N_r t_r+N_k t_k
\label{time1}
\end{equation}
with $N_r,\;N_k$ being the numbers of terms in each sum. Every
one of these sums converges when $N_r,\;N_k\rightarrow \infty$ to a certain
value, depending on $\alpha$, while the sum of the limiting values is a
constant. We can take into account the errors, corresponding to each of the
sums separately. For $\alpha\rightarrow 0$ the error for the real space term
is zero and the other one tends to infinity (and vice versa as
$\alpha\rightarrow \infty$). The minimum total error should therefore
correspond to the value of $\alpha$, satisfying the relation
$\di (\delta E_r + \delta E_k)/\di \alpha = 0$.

Focusing on the 2D system of our example, we note
that the long-range expansions of the terms in (\ref{defrho}) and
(\ref{defkappa2d})  are
similar, in a sense that the leading terms in both expressions are
Gaussians,
\begin{eqnarray}
\frac{\Gamma(k/2,\alpha^2n^2)}{\Gamma(k/2)n^k} & = & \exp(-\alpha^2n^2)\left[\frac{C_r}{n^2}+
O\left(\frac{1}{|\bs{n}|^3}\right)\right] \label{3drealtermexpand}\\
\frac{\pi^{\frac{3}{2}}\alpha^{k-3}}{\Gamma(k/2)}E_{\frac{k-1}{2}}\left(\frac{\pi^2n^2}{\alpha^2}
\right) & = & \exp\left(-\frac{\pi^2n^2}{\alpha^2}\right)\left[\frac{C_k}{n^2}+
O\left(\frac{1}{|\bs{n}|^3}\right)\right] \ .
\label{3dmomtermexpand}
\end{eqnarray}
The power-law terms in $\bs{n}$ and the constants $C_r,\:C_k$ may be
neglected since the leading behaviour is driven by the Gaussian. The cut-off
errors due to finite numbers of elements in the sums can be evaluated by
ignoring the discrete structure of the images and approximating the sums by
uniform integrals,
\begin{eqnarray}
\delta E& =& \int_R^{\infty}\exp(-\alpha^2 r^2) 2\pi r \di r +
\int_K^{\infty}\exp\left(-\frac{\pi^2k^2}{\alpha^2}\right) 2\pi k \di k \nonumber \\
& = & \pi\left[\frac{\exp(-\alpha^2 N_r/\pi)}{\alpha^2}+
\frac{\alpha^2 \exp\left(- \pi N_k/\alpha^2\right)}  {\pi^2}\right]
\label{dE1}
\end{eqnarray}
with $R\simeq \sqrt{N_r/\pi}$ and $K\simeq \sqrt{N_k/\pi}$ the
approximate cut-off lengths in  real and momentum spaces, respectively.
The optimal value for $\alpha$ can be obtained by solving the equation $\di \delta E_r/\di \alpha = -\di \delta E_r/\di \alpha $. The first-order approximation of this equation is found by taking logarithms of both sides and omitting constants and terms, depending on $\alpha$ logarithmically, that is
\begin{equation}
A_k/\alpha^2-A_r \alpha^2= 0
\label{dE2}
\end{equation}
with $A_k=\pi N_k$ and $A_r=N_r/\pi$, which yields
\begin{equation}
\alpha=(A_k/A_r)^{1/4}=\left(\pi^2 N_k/N_r\right)^{1/4} \ .
\end{equation}
Then, at lowest order one finds~(\ref{dE1}),
\begin{equation}
\delta E\sim  \exp(-\alpha^2 N_r/\pi)=\exp(-\sqrt{N_k N_r}) \ .
\end{equation}
Since the calculation time is linear with the numbers of elements $N_r$ and
$N_k$, we may conclude that with $N_k$ fixed and comparatively large $N_r$,
$\;\ln(\delta E) \sim \sqrt{N_r} \sim \sqrt{T}$ and vice versa, with
$N_r$ fixed and large $N_k$, $\ln(\delta E) \sim \sqrt{N_k} \sim \sqrt{T}$.
This power law may be easily checked in our calculations, as it is shown in
Fig.~(\ref{erront4}). Note that for the obtained value of $\alpha$ the errors of the real- and momentum-space cutoffs are of the same order of magnitude, that is
$\delta E_r \approx \delta E_k$, which may serve as a rough criterium to optimise the parameter $\alpha$.

A more advanced procedure for optimisation of the parameters, proposed by
Perram \textit{et al.}~\cite{Perram1988}, yields an asymptotic scaling
$N^{3/2}$, with $N$ the number of particles. It is based on the form of Ewald summation with the momentum space
sum, linear in $N$ (\ref{pigeneff}). Suppose the values of the calculation
time $t_r$, $t_k$ to perform unit computations in both sums are known and
the target error level $\exp(-p)$ is fixed. Then, the total execution time
in the real and momentum spaces is
\begin{equation}
T=T_r+T_k=N^2\pi R^2 t_r+N \pi K^2 t_k
\end{equation}
with $p=\alpha^2 R^2=\pi^2 K^2/\alpha^2$. Expressing $K$ as $K=p/(\pi R)$
we can see that the minimum of the total time T corresponds to
\begin{eqnarray}
R_{\rm opt}=&&\left(\frac{p}{\pi}\right)^{1/2}\left(\frac{t_k}{t_r}\right)^{1/4}N^{-1/4}\\
K_{\rm opt}=&&\left(\frac{p}{\pi}\right)^{1/2}\left(\frac{t_k}{t_r}\right)^{-1/4}N^{1/4}\\
\alpha_{\rm opt}=&&\sqrt{\pi}\left(\frac{t_k}{t_r}\right)^{-1/4}N^{1/4} \ .
\end{eqnarray}
The computation time is equally divided between the real and
momentum space parts (this was also stated in our simple optimisation
scheme), with a scaling of the whole summation given by
\begin{equation}
T=2 N^2 \pi R^2 t_r= 2 p\sqrt{t_r t_k}N^{3/2}
\end{equation}
Notice that the values of the free parameters change very slowly when
the simulation cell is enlarged, and in particular $\alpha$ is not affected
by the choice of the precision. Similar formulae for the optimised
parameters in three-dimensional systems, with a discussion of different
techniques to improve performance of the Ewald summation, are given by
Fincham~\cite{Fincham1994}. A more precise and detailed analytic study
of the cut-off errors with verifications of the analytic results in actual
calculations can be found in the work of Kolafa and Perram~\cite{Kolafa1992}.
An optimised method for treating the truncation error in Ewald sums with
generic potentials was proposed by  Natoli and
Ceperley~\cite{Natoli1995}. While the needed CPU time scales as $\mathcal{O}
(N \ln N)^{3/2}$, it was shown that in the example of the Coulomb potential the
method resulted in greatly improved accuracy compared to that of standard
Ewald technique for a comparable computational effort.
This method is based on an expansion of the real
space function in an arbitrary radial basis with a parametric set of
numbers in place of the $k$-dependent prefactors of $\exp(2\pi\ii
\bs{n}\bs{r})$. The subsequent minimization of $\chi^2$ with respect to the
whole set of parameters yields a final optimal solution, that is the real
space expansion coefficients and the $k$-space factors.  This technique
was also applied to derive the optimised summation formulae for the
two-dimensional Coulomb system~\cite{Holtzmann2005}.

In general, the unit computation time in momentum space is 2--4 times faster
than the one in real space. Taking the following reasonable assumptions
$p=4\pi$, $t_k/t_r=3$, we find $R_{\rm opt}\approx 2.6/N^{1/4}$. We want
$R$ to be below 0.5, since in this case the summation in the real space
reduces to the accumulation of the single component $\bs{n}=0$. This
condition $R_{\rm opt}=0.5$, with our previous assumptions, corresponds to
\begin{equation}
N_{\rm opt}= 770,\:\:K_{\rm opt}=8.0,\:\:\alpha_{\rm opt}=7.1
\end{equation}
In smaller systems, the other components of the real sum, starting from
$|\bs{n}|=1$, should be considered.

It is worth pointing out that if the interaction is very
strong at short distances (as for the Lennard-Jones potential), then
in principle the real-space cut-off $R$ can be chosen below the ``hard core
radius'' with a large enough value of $\alpha$. This leads to the possibility
of dropping completely the real-space part of the total sum  and treat the
$k$-space only. This can be advantageous in different aspects, especially
with the current progress in the development of efficient FFT-based
methods. Nonetheless, we are not aware of any present
application of a similar technique.

\section{Conclusions}
\label{sec:conclusions}

In the present work, we have applied the Ewald summation method to
$1/|\bs{r}|^k$ polytropic potentials in three-, two- and
one-dimensional geometries in a simulation box with periodic boundary
conditions. We have found the explicit functional forms for all the
components of the sums in both real and momentum spaces, with
special attention being paid to the cases of long-range interactions, that
is conditionally convergent or divergent potentials (corresponding to
$k<D$, with $D$ standing for the dimensionality), ``marginal''
interactions ($k=D$), and short-range interactions (with $k>D$). For the
latter case of short-range interaction potentials, where in principle a
straightforward summation of the initial sum (\ref{initialhamiltonian1}) is
possible, the Ewald method is shown to be useful, as it yields the faster
(Gaussian) convergence rate. A condition of charge neutrality of the simulation cell is
stated to be necessary for conditionally convergent and divergent
potentials; a homogeneous positive charge background (``jellium'' model) is
introduced as the most relevant and frequently used kind of neutralization.
The conditionality of the convergence for a charge-neutral system, governed
by the Coulomb interaction, is discussed with a justification of the use of
a specific periodicity-preserving convergence factor. The derivation
technique, presented in our work, is consistent with the arguments of
de Leeuw \textit{et al.}~\cite{leeuw}.

The results are first presented for the case of a 3D system
in a cubic simulation box in order to explain the
general mathematical procedure, which for the specific case of the Coulomb
potential recovers well-known results~\cite{allen}. Later on, the same
mathematical technique is applied to 2D and 1D geometries.
For the one-dimensional case the
initial sum for the potential energy  is explicitly evaluated (\ref{ep1ddirect2}),
nonetheless the Ewald summation is developed for this
case too and may be used as a mathematical equality. The special
representations of the reciprocal space sums, linear in the number of
particles $N$ and hence more efficient in actual modeling, are presented
for 3D and 2D systems. The explicit expressions for the terms of the Ewald
sums are given in a tabular form for physically relevant potentials
with small integer power indexes $k$, as dipole-dipole interaction
potential, Lennard-Jones potential and others in both three- and
two-dimensional geometries 
(see Tables~1 and 2).

When the simulation box cannot be chosen cubic, for example in a modeling
of a three-dimensional hcp crystal structure, the Ewald method can also be
applicable after a certain modifications. Formally, it consists in the choice
of an appropriate rectangular simulation box and a substitution of the
vector $\bs{n}$ by $\bs{n}_r=(\bs{n}_xL_x+\bs{n}_yL_y+\bs{n}_zL_z)/L_0$ and
$\bs{n}_k=(\bs{n}_x /L_x+\bs{n}_y /L_y+\bs{n}_z /L_z)L_0$ in the real
and momentum space sums, respectively [see
(\ref{i01infpolytropicnbox}) and (\ref{3djacobinbox})].

The optimisation of the involved parameters, that is cut-off numbers in
both sums and the integration parameter $\alpha$, is a necessary operation
in order to improve the convergence rates and avoid excessive calculations.
The main idea of the optimisation, proposed in the present work, is to
perform a benchmark calculation, minimizing the variance of the result. A
particular example of the application of the technique is presented for a
calculation of the potential energy of a two-dimensional gas of dipoles, aligned
perpendicular to the plane of motion. This practical optimisation technique
is thought to be efficient for stationary and nearly uniform systems that
appear, for instance, in Monte Carlo simulations. In spite of being very
simple, it allows to find rather quickly adequate parameter ranges. The
analytical estimations of the parameters are given as well and are proven
to be consistent with the results, obtained in our method. A more
sophisticated method to optimise the calculation parameters, taking
advantage of the $\mathcal{O}(N)$ representation of the Fourier transform
sum, is also presented with explicit estimations of the parameters for a
typical system simulated by Quantum Monte Carlo methods.

\section*{Acknowledgements}
We acknowledge partial financial support by DGI
(Spain) under Grant No. FIS2008-04403 and Generalitat de Catalunya under
Grant No. 2009-SGR1003.

\section*{Appendix}
We prove that the sums $S_{-+}$ and $S_{++}$
(\ref{psisum2}--\ref{psisum3}) vanish on average, allowing to calculate the
potential energy over the negatively charged particles' positions only.
\begin{itemize}
\item
First, let us show that the integral of $\psi$ over the cell is zero.
Since the distances are in the units of $L$, consider the cubic cell
$\Omega = (x,y,z) \in [-1/2,1/2]^3$, that yields
\begin{equation}
\int_{\Omega}\psi(\bs{r}) \di \bs{r} = J_1 + J_2 + C_1
\label{intpsi1}
\end{equation}
where
\begin{eqnarray}
J_1 &&= \int_{\Omega}\di \bs{r} \sum_{\bs{n}}R(\bs{n},\bs{r})\label{j11} \\
J_2 &&= \int_{\Omega}\di \bs{r} \sum_{\bs{n}\neq \bs{0}}K(\bs{n},\bs{r}) \
.
\label{j21}
\end{eqnarray}
It can be easily seen, that the second integral $J_2$ is zero,
\begin{eqnarray}
J_2   &&= \sum_{\bs{n}\neq
\bs{0}}\kappa(\bs{n},\bs{r})\int_{\Omega}\cos(2 \pi \bs{n}
\bs{r})\di \bs{r}\nonumber \\
&&= \sum_{\bs{n}\neq \bs{0}}\kappa(\bs{n}) \frac{\sin(2 \pi (n_x+n_z+n_z))}
{(2 \pi)^3 n_x n_y n_z} = 0
\end{eqnarray}

As far as the integral $J_1$ is concerned, we can notice that the
regions $\Omega'(\bs{n})=\bs{r}+\bs{n}$, where $\bs{r} \in
\Omega,\;\bs{n} \in \mz^3$
are the same cubic unit cells, displaced by an integer vector, thus
covering all
the coordinate space with only zero-measure intersections.
It means that the summation of the integrals in (\ref{j11}) over the
cell $\Omega$
can be substituted by the integration over the whole coordinate space,
\begin{eqnarray}
J_1 &&= \sum_{\bs{n}} \int_{\Omega'(\bs{n})}R(\bs{n},\bs{r})\di \bs{r}=
\sum_{\bs{n}} \int_{\Omega'(\bs{n})}\frac{\Gamma (\frac{k}{2},\alpha
^2|\bs{r}+\bs{n}|^2 )}
{\Gamma (\frac{k}{2})|\bs{r}+\bs{n}|^k}\nonumber \\
&&= \alpha^{k-3} \int_{\mr^3 \setminus 0}\frac{\Gamma
(\frac{k}{2},\bs{\rho}^2)}
{\Gamma (\frac{k}{2})\bs{\rho}^k}\di \bs{\rho} = -\frac{2 \pi
^{\frac{3}{2}}
\alpha ^{k-3}}{(k-3) \Gamma\left[\frac{k}{2}\right]}=-C_1 \ ,
\end{eqnarray}
and thus the whole integral (\ref{intpsi1}) is equal to zero.

\item
Consider two species of the particles: negative charges $q_i$ on
positions $\bs{r}_i$ and a positively charged and uniformly distributed background
with a total charge $q_{+}N_{+}=-q_i N_i$, ensuring the neutrality of the cell.
 Let us demonstrate that $S_{-+}$ is equal to zero, when the number of background charges tends to infinity.
In this case the sum (\ref{psisum2}) for $S_{-+}$ may be rewritten as an integral over
 the background charges' positions
\begin{equation}
S_{-+}=\sum_i q_i \int_{\Omega}\psi(\bs{r}_p-\bs{r}_i)\sigma \di
\bs{r}_p = \sum_i q_i \int_{\Omega_i}\psi(\bs{r})\sigma \di  \bs{r} \ ,
\label{Sepapp1}
\end{equation}
where we did the change of variables $\bs{r}=\bs{r}_p-\bs{r}_i$. The
regions $\Omega$ and $\Omega_i$ refer to the original simulation cell
and the same cell, moved by the vector $\bs{r}_i$,
 and $\sigma$ stands for the background charge density $\sigma=-q_i N_i/V(\Omega)$. It is clear that
every vector $\bs{r}=(x,y,z)\in \Omega_i$ can be displaced into the cell
$\Omega$ by the corresponding shift
$\tilde{\bs{r}}=(\tilde{x},\tilde{y},\tilde{z})=(x-aL,y-bL,z-cL)\in
\Omega$ with integers $a,\:b,\:c$. The Jacobian $J$ of the change of
variables $\bs{r}\rightarrow\tilde{\bs{r}}$ is obviously 1. On the other
hand, due to the periodicity of $\psi$,
\begin{equation}
\psi(\bs{r}) = \psi(\tilde{\bs{r}})\ ,
\end{equation}
and $\tilde{\bs{r}}$ runs over the whole region $\Omega$ due to the
conservation of the volume with $J=1$. Finally, Eq.~(\ref{Sepapp1}) can
be written as
\begin{equation}
S_{-+}=\sum_i q_i \int_{\Omega}\psi(\tilde{\bs{r}}) \sigma \di
\tilde{\bs{r}} = 0\ .
\end{equation}

In the similar manner,
 the interaction between the charges of the background $S_{++}$ in the limit $N_{+}\rightarrow\infty$ is
given by the double integral
\begin{equation}
S_{++}=\frac{1}{2}\int_{\Omega}\di \bs{r}_1\int_{\Omega} \di \bs{r}_2
\psi(\bs{r}_1-\bs{r}_2) \sigma^2  =0 \ ,
\label{Sppapp1}
\end{equation}
since $\int_{\Omega}\psi(\bs{r}_1-\bs{r}_2) \di \bs{r}_2=0$, following
the same arguments as for the case of $S_{-+}$.
\end{itemize}

\end{document}